\documentclass[12pt]{article}
\usepackage{amsmath,amssymb}
\usepackage{amsthm} 
\usepackage{lmodern}
\usepackage{iftex}
\usepackage{indentfirst}
\usepackage{setspace}
\usepackage{enumitem} 
\usepackage{graphicx}
\usepackage{listings}
\usepackage{float} 
\usepackage{subfigure}
\usepackage[numbers,sort&compress]{natbib}
\usepackage[ruled,linesnumbered]{algorithm2e}
\usepackage{bm}
\usepackage{color}
\usepackage{multirow}
\usepackage{makecell}
\usepackage{booktabs}
\usepackage{boxedminipage}  
\usepackage{natbib}
\usepackage{xcolor}
\setcitestyle{authoryear,round}
\usepackage[colorlinks=true,linkcolor=red,citecolor=blue,urlcolor=black]{hyperref}
\usepackage[title]{appendix}
\newtheorem{theorem}{Theorem}
\newtheorem{corollary}{Corollary}
\newtheorem{lemma}{Lemma}
\newtheorem{definition}{Definition}
\newtheorem{assumption}{Assumption}
\newtheorem{remark}{Remark}

\newtheorem{proposition}{Proposition}

\newcommand{\huEdit}[1]{\textcolor{black}{#1}}

\title{\vspace{-1.5cm}Policy Optimization and Multi-Agent Reinforcement Learning for Mean-Variance Team Stochastic Games }
\author{Junkai Hu\thanks{J. Hu is with the School of Mechanical and Electrical Engineering, Shenzhen Polytechnic University, Shenzhen
518055, China.}, \quad Li Xia\thanks{ L. Xia is with the School of Business, Sun Yat-Sen University, Guangzhou 510275, China. (email: xiali5@sysu.edu.cn)}}
\date{}

\begin{document}

\maketitle
\vspace{-0.2cm}
\begin{abstract}
    We study a long-run mean-variance team stochastic game (MV-TSG), where agents act independently to optimize a shared mean-variance objective. MV-TSG presents two key challenges: (1) the variance metric's non-additivity and non-Markovian nature in dynamic settings, and (2) non-stationary environments due to simultaneous policy updates of all agents. Both challenges render dynamic programming inapplicable. In this paper, we propose a novel Mean-Variance Multi-Agent Policy Iteration (MV-MAPI) algorithm for MV-TSGs based on the sensitivity-based optimization theory and a sequential update scheme. We prove that MV-MAPI converges monotonically to a first-order stationary point, and derive conditions under which such points correspond to (local) Nash equilibria or even strict local optima. We further develop a multi-agent reinforcement learning algorithm based on MV-MAPI. Numerical experiments on energy management in multiple microgrid systems validate our main results. To the best of our knowledge, this is the first work to propose theoretically provable algorithms for MV-TSGs.
    
\end{abstract}


\section{Introduction}
Stochastic games, proposed by Nobel laureate \cite{shapley1953stochastic} and also known as Markov games \citep{littman1994markov}, integrate matrix games with Markov chains to provide a general framework for modeling sequential decision-making problems involving multiple agents (or decision makers). In such games, agents act \textit{independently} based on state-dependent policies, and the game state evolves according to a transition probability matrix determined by the joint actions of all agents. At each time step, agents receive immediate rewards as a consequence of both the current state and the joint action taken. Each agent seeks to maximize its cumulative reward—either discounted or undiscounted—over a given time horizon.

According to the reward relationship among agents, stochastic games can be classified into three groups: team stochastic games (TSGs, also known as cooperative Markov games \citep{zhong2024heterogeneous}), zero-sum stochastic games, and general-sum stochastic games. These categories correspond to cooperative, competitive, and mixed settings in the field of multi-agent reinforcement learning (MARL) \citep{yang2020overview}. The most widely used solution concept in stochastic games is the Nash equilibrium (NE), which characterizes a strategy profile where no agent can improve its expected return by unilaterally deviating from its current policy, given that the policies of all other agents are fixed.

TSGs, a subclass of Markov potential games \citep{leonardos2022global, zhang2024gradient}, model cooperative multi-agent systems with a common reward function. Early works, such as \cite{marschak1955elements} and \cite{radner1962team}, study single-stage cooperative decision-making, which was later extended to dynamic settings, thereby contributing to the development of team theory \citep{marschak1972economic,ho1980team}. Practical applications include supply chain coordination \citep{oroojlooyjadid2022deep}, resource balancing in logistics \citep{li2019cooperative}, and multi-order execution in finance \citep{fang2023learning}. The prevalence of team collaboration in practice and the remarkable success of artificial intelligence, particularly reinforcement learning (RL), have spurred substantial attention to TSGs and cooperative MARL.

A straightforward approach for addressing TSGs is to have each agent independently update its policy using single-agent methods. However, this often leads to a non-stationarity problem, as simultaneous policy updates by multiple agents change the environment dynamics from the perspective of each individual agent. This violates the stationary environment assumption required by single-agent algorithms, making their direct application to TSGs ineffective.

MARL provides a framework for approximately solving TSGs, particularly in scenarios where the environmental model parameters are unknown. In this setting, each agent updates its policy based on data collected through interactions with the environment. Early cooperative MARL algorithms, such as Team-Q \citep{littman2001value}, Distributed-Q \citep{lauer2000algorithm}, JAL (Joint Action Learner) \citep{claus1998dynamics}, and OAL (Optimal Adaptive Learning) \citep{wang2002reinforcement}, were developed for specific problem settings. However, their applicability is often limited by strong assumptions, including the presence of a centralized decision-maker, deterministic state transitions, or single-stage formulations. Recent decentralized algorithms, such as those proposed by \cite{arslan2016decentralized} and \cite{yongacoglu2021decentralized}, focus on discounted TSGs and are shown to converge—almost surely—to NEs or optimal NEs (maximize the team’s objective). Their reliance on random exploration and pairwise comparisons often leads to low sample efficiency and slow convergence, limiting scalability in complex environments.

Over the past decade, the centralized training with decentralized execution (CTDE) paradigm \citep{kraemer2016multi} has become a prevalent MARL framework. In the decentralized execution phase, agents interact with the environment online by taking actions according to their individual policies. In contrast, during the offline centralized training phase, complete trajectory data from all agents are accessible and utilized for policy updates. 

CTDE-based cooperative MARL algorithms can generally be categorized into value decomposition methods and policy gradient methods. Value-decomposition approaches aim to factorize the joint action-value function $Q^{tot}$ into individual action-value functions  $Q^{ind}_i$ for each agent $i$. A key requirement for the effectiveness of these methods is the Individual-Global-Maximum (IGM) condition \citep{sunehag2018value}, given by: 
\begin{equation*}
    \mathop{\arg\max}_{\bm{a}}Q^{tot}(s,\bm{a})= 
    \begin{pmatrix}
    \mathop{\arg\max}\limits_{a_1}Q^{ind}_1(s,a_1) \\
    \vdots \\
    \mathop{\arg\max}\limits_{a_N}Q^{ind}_N(s,a_N)
    \end{pmatrix} 
    \qquad (\text{IGM condition}),
    \notag
\end{equation*}
where $N$ is the number of agents, $a_i$ is the action taken by agent $i$, and $\bm{a}=(a_1,\ldots,a_N)$ denotes the joint action. 
However, finding individual action-value functions  $Q^{ind}_i$ that satisfy the IGM condition is challenging. Value-decomposition algorithms typically employ specifically designed neural networks to approximate these functions \citep{rashid2020monotonic}, thereby lacking rigorous theoretical guarantees.
Policy-gradient methods, originally developed for single-agent Markov decision processes (MDPs), have also been extended to multi-agent settings. Nevertheless, these methods are affected by the environmental non-stationarity arising from concurrent policy updates of multiple agents \citep{kuba2022trust}. For more details on state-of-the-art policy gradient-based MARL algorithms, please refer to \cite{foerster2018counterfactual,yu2022surprising,zhong2024heterogeneous}. It should be noted that these works mentioned above are primarily concerned with risk-neutral settings with the discounted accumulated reward.

\huEdit{Risk preferences are crucial in decision-making, particularly in stochastic environments \citep{cavazos2023average}. Variance, a key measure of reward variability, is widely used to characterize risk, as in stochastic congestion games \citep{nikolova2014mean,lianeas2019risk} and dual-sourcing problems \citep{gupta2020dual}. \cite{slumbers2023game} study risk-averse equilibria where each agent seeks to minimize the variance caused by others’ actions; however, their algorithms provide no convergence guarantees. Despite these contributions, most existing studies focus on static or normal-form games, leaving risk-aware analysis of stochastic dynamic games largely unexplored.}

Only limited research has investigated stochastic games with risk-sensitive objectives. Some studies establish the existence of equilibria under specific risk measures, such as exponential utility \citep{bauerle2017zero} and conditional value at risk (CVaR) \citep{liu2023zero}, without proposing corresponding solution algorithms. \cite{etesami2018stochastic} and \cite{wu2024relationships} introduce heuristics for prospect-theoretic stochastic games, but their algorithms may traverse all possible policies in the worst case, leading to prohibitive computational complexity. More recently, \cite{mazumdar2025tractable} provide convergence guarantees for risk-averse quantal response equilibria in finite-horizon stochastic games, marking a promising step toward risk-sensitive equilibrium computation.

As evidenced by the above literature, solving stochastic games under risk-sensitive objectives has attracted growing attention but remains a significant challenge. Research on risk-averse TSGs or cooperative MARL is still limited.  Recent works \cite{qiu2021rmix} and \cite{shen2023riskq} investigate cooperative MARL with risk-sensitive utilities, but rely on value decomposition and lack rigorous theoretical analysis. To the best of our knowledge, no existing work provides algorithms with theoretical guarantees for risk-sensitive TSGs or cooperative MARL. 

In this paper, we investigate mean-variance team stochastic games (MV-TSGs), where agents seek policies that collectively maximize the long-run mean-variance of common rewards. Addressing MV-TSGs poses two fundamental challenges: (1) the non-additive and non-Markovian nature of the variance metric, as it depends on both current and future joint actions in a dynamic setting; and (2) environmental non-stationarity, since each agent’s policy forms part of the environment for other agents, and simultaneous policy updates induce non-stationarity from each agent’s perspective. These challenges prevent the direct application of classical dynamic programming techniques.

\huEdit{To solve MV-TSGs, we first provide the optimization direction of the joint policy using sensitivity-based optimization theory. Based on this, we propose a Mean-Variance Multi-Agent Policy Iteration (MV-MAPI) method by introducing a sequential update scheme for individual policy updates. We prove that MV-MAPI converges monotonically to a first-order stationary point, and derive conditions under which such points are (local) Nash equilibria or strict local optima. Moreover, we develop a modified MV-MAPI that escapes undesirable stationary points and converges to a strict local optimum. To handle large-scale MV-TSGs with unknown environmental parameters, we extend trust region optimization to the mean-variance setting and propose an MARL algorithm, Mean-Variance Multi-Agent Trust Region Policy Optimization (MV-MATRPO), following the framework of \cite{zhong2024heterogeneous}. A performance lower bound is established for each joint policy update, guaranteeing monotonic improvement when the trust region is sufficiently tight. The effectiveness of our algorithms is demonstrated by numerical experiments on an energy management problem involving multiple microgrid systems (MMSs).}

\huEdit{The contributions of this paper are twofold. First, we propose MV-MAPI and demonstrate that, unlike in discounted or average-reward TSGs where first-order stationary points coincide with NEs \citep{cheng2024provable,zhang2024gradient}, the local geometry of these points in MV-TSGs exhibits distinct characteristics. We establish verifiable conditions under which such points correspond to (local) NEs or even strict local optima, and propose a modified MV-MAPI guaranteed to converge to strict local optima. Second, we propose MV-MATRPO and derive a performance lower bound for policy updates. Compared with existing relevant works by \cite{qiu2021rmix} and \cite{shen2023riskq}, to the best of our knowledge, this is the first work to develop algorithms with theoretical guarantees for risk-sensitive cooperative MARL.}


The remainder of this paper is organized as follows. Section~\ref{sec:preliminary} introduces the MV-TSG problem. Section~\ref{sec:MAPI} presents the MV-MAPI algorithm and analyzes its convergence properties. Subsequently, the MV-MATRPO algorithm is developed in Section~\ref{sec:MV-MARL}.  Section~\ref{sec:exp_results} validates the effectiveness of our algorithms through an energy management problem of MMSs. Finally, Section~\ref{sec:conclusion} concludes this paper.

\subsection{Notations}
\huEdit{Let $X := \{x_1, \ldots, x_n\}$ be a finite set.  
We use $\bm{x}_{-i}$ to denote the set of all elements in $X$ except $x_i$.  
Let $i_{1:h}$ represent an $h$-element ordered subset of $X$, and let $-i_{1:h}$ denote its complement in $X$.  
The symbol $i_k$ refers to the $k$-th element of $i_{1:h}$.  
For a finite set $Y$, let $\Delta(Y)$ denote the set of probability distributions over $Y$.}

\section{Problem Setting}
\label{sec:preliminary}

We consider an infinite-horizon discrete-time TSG, denoted by $\langle\mathcal{N},\mathcal{S}, \mathcal{A}, P, r \rangle$. Here, $\mathcal{N}=\{1,\ldots, N\}$ is the finite set of agents, $\mathcal{S}$ is the finite system state space, $\mathcal{A}=\prod_{i\in \mathcal{N}}\mathcal{A}_i$ is the finite joint action space, $\mathcal{A}_i$ is the action space of agent $i\in \mathcal{N}$,  $P:\mathcal{S} \times \mathcal{A} \mapsto \Delta(\mathcal{S})$ is the state transition probability function, 
and $r:\mathcal{S} \times \mathcal{A} \mapsto \mathbb{R}$ is the common reward function shared by all agents. Each agent $i \in \mathcal{N}$ follows a stationary policy  $\mu_i:\mathcal{S} \mapsto \Delta(\mathcal{A}_i)$, with policy space $\mathcal{U}_i$. 
The joint policy is $\bm\mu=(\mu_1,\ldots,\mu_N)$, and the joint policy space is $\mathcal{U} = \prod_{i\in \mathcal{N}}(\mathcal{U}_i)$. 
For a given state $s$, the probability that agents choose joint action $\bm{a}=(a_1,\ldots,a_N)$ is $\bm\mu(\bm{a}|s)=\prod_{i\in\mathcal{N}}\mu_i(a_i|s)$. If $\bm\mu$ is  deterministic, i.e., $\bm\mu:\mathcal{S}\mapsto\mathcal{A}$, we call it a deterministic joint policy. Let $\mathcal{D}$ denote the deterministic joint policy space and we have $\mathcal{D} \subset \mathcal{U}$.
We make the following ordinary ergodic assumption in this study.

\begin{assumption}\label{assum1}
The Markov chain induced by any joint policy $\bm\mu \in \mathcal{U}$ is ergodic.
\end{assumption}

At each time step $t$, each agent $i$ adopts an action $a_{i,t}$ according to the system state $s_t$ and its policy $\mu_i$. With the joint action $\bm{a}_t$, the system will transit to the next state $s_{t+1}$ with transition probability function $P(s_{t+1}|s_t,\bm{a}_t)$ and an immediate common reward $r(s_t,\bm{a}_t)$ will be incurred. We define the long-run average reward of TSGs under a joint policy $\bm\mu$ as 
\begin{equation*}
    \eta^{\bm{\mu}}:=\lim\limits_{T \rightarrow \infty} \frac{1}{T} \mathbb{E}_{\bm{\mu}} \Big[ \sum\limits_{t=0}^{T-1}r(s_t,\bm{a}_t)  \Big], \notag
\end{equation*}
where $\mathbb{E}_{\bm{\mu}}$ stands for the expectation with $\bm{a}_t \sim \bm{\mu} (\cdot|s_t)$, $s_{t+1} \sim P(\cdot | s_t,\bm{a}_t)$. When $T \rightarrow \infty$, $\eta^{\bm{\mu}}$ is independent of the initial state $s_0$, with ergodic Assumption~\ref{assum1}. We denote $\pi^{\bm\mu}: \mathcal{S} \mapsto \Delta(\mathcal{S})$ as the steady state distribution under the joint policy $\bm\mu$, and Assumption~\ref{assum1} ensures that $\pi^{\bm\mu}(s)>0$ for any joint policy $\bm\mu$ and state $s$. Then, the long-run average reward can be rephrased as 
\begin{equation}
    \eta^{\bm{\mu}}= \mathbb{E}_{s\sim \pi^{\bm{\mu}}, \bm{a} \sim \bm{\mu}}[r(s,\bm{a})]  
    = \sum\limits_{s \in \mathcal{S}} \pi^{\bm\mu}(s) \sum\limits_{\bm{a} \in \mathcal{A}} r(s,\bm{a})\bm\mu(\bm{a}|s).  \label{eq:eta-2} 
\end{equation}

Additionally, we are concerned with the long-run variance of the common reward, which describes the variability of the steady reward distribution and has been studied in single-agent settings by \cite{filar1989variance}, \cite{sobel1994mean} and \cite{xia2025global}. For TSGs, the long-run variance is defined as
\begin{equation*} 
    \zeta^{\bm{\mu}} := \lim\limits_{T \rightarrow \infty} \frac{1}{T} \mathbb{E}_{\bm{\mu}} \Big[ \sum\limits_{t=0}^{T-1}(r(s_t,\bm{a}_t) - \eta^{\bm{\mu}} )^2   \Big]. 
\end{equation*}
\huEdit{In MV-TSGs, the objective is to optimize the following mean-variance performance metric}
\begin{align}\label{eq:mean-variance}
     J^{\bm{\mu}} &:= \lim\limits_{T \rightarrow \infty} \frac{1}{T} \mathbb{E}_{\bm{\mu}} \Big\{ \sum\limits_{t=0}^{T-1} [r(s_t,\bm{a}_t) - \beta(r(s_t,\bm{a}_t) - \eta^{\bm{\mu}} )^2]   \Big\} \\
     &=  \eta^{\bm{\mu}} - \beta \zeta^{\bm{\mu}}, \notag
\end{align}
where $\beta \ge 0$ is the parameter for the trade-off between mean and variance. For notational convenience, we denote $J^{\bm\mu}$ and $\eta^{\bm\mu}$ by $J(\bm\mu)$ and $\eta(\bm\mu)$, respectively, when necessary.

Inspired by (\ref{eq:mean-variance}), we define the surrogate reward function associated with the mean-variance metric as  
\begin{align}
    f^{\bm\mu}(s,\bm{a})=r(s,\bm{a}) - \beta(r(s,\bm{a})-\eta^{\bm{\mu}})^2.
    \label{eq:reward_f}
\end{align} 
Similarly to (\ref{eq:eta-2}), the mean-variance performance function can be computed by 
\begin{equation}
    J^{\bm{\mu}}=\sum\limits_{s \in \mathcal{S}} \pi^{\bm\mu}(s) \sum\limits_{\bm{a} \in \mathcal{A}} f^{\bm\mu}(s,\bm{a})\bm\mu(\bm{a}|s) 
    \label{eq:eta_f}.
\end{equation}

According to the definition of the value function in the average-reward setting \citep{sutton2018reinforcement},\footnote{\huEdit{Unlike the discounted case, the value function in the average-reward setting, also called the average-reward bias function, is defined as $V^{\bm\mu}(s):= \mathbb{E}_{\bm{\mu}} \Big[\sum\limits_{t=0}^\infty (r(s,\bm{a})- \eta^{\bm\mu})|s_0 = s \Big]$, which characterizes the cumulative advantage of system rewards relative to the average reward $\eta^{\bm\mu}$ starting from state $s$.}} we define the value function $V_{f}^{\bm{\mu}}$, the action-value function $Q_{f}^{\bm{\mu}}$, and the advantage function $A_{f}^{\bm{\mu}}$ in MV-TSGs with respect to the surrogate reward function $f^{\bm{\mu}}$ as follows:
\begin{align} 
    V_{f}^{\bm{\mu}}(s)
    :&= \mathbb{E}_{\bm{\mu}} \Big[\sum\limits_{t=0}^\infty (f^{\bm\mu}(s_t,\bm{a}_t)- J^{\bm\mu})|s_0 = s \Big], \notag\\
    Q_{f}^{\bm{\mu}}(s,\bm{a}):&=\mathbb{E}_{\bm{\mu}} \Bigl[
        \sum\limits_{t=0}^\infty 
        (f^{\bm\mu}(s_t,\bm{a}_t) -J^{\bm\mu})|s_0=s,\bm{a}_0=\bm{a}
    \Bigr] \notag\\
    & = f^{\bm\mu}(s,\bm{a})  -J^{\bm\mu} + \sum\limits_{s' 
    \in \mathcal{S}}P(s'|s,\bm{a})V_{f}^{\bm\mu}(s'),  \label{eq:qf function-2 }\\
    \label{eq:af function} A_{f}^{\bm{\mu}}(s,\bm{a}):&= Q_{f}^{\bm{\mu}}(s,\bm{a}) - V_{f}^{\bm{\mu}}(s).
\end{align}
The strong Markov property indicates that the value function  satisfies the Poisson equation  
\begin{equation*}\label{eq:bellman equation}
    V_{f}^{\bm{\mu}} (s) = f^{\bm{\mu}}(s) - J^{\bm{\mu}} + \sum\limits_{s' 
    \in \mathcal{S}}P^{\bm{\mu}} (s'|s)V_{f}^{\bm{\mu}} (s') \quad \forall s \in \mathcal{S},
\end{equation*}
where $P^{\bm\mu}(s'|s)=\sum\limits_{\bm{a}\in \mathcal{A}}\bm\mu(\bm{a}|s)P(s'|s,\bm{a})$, $f^{\bm{\mu}}(s)=\sum\limits_{\bm{a} \in \mathcal{A}} \bm{\mu}(\bm{a}|s)f^{\bm\mu}(s,\bm{a})$. It is known that $P^{\bm\mu}$ is a stochastic matrix and its rank is $|\mathcal{S}|-1$. 

\huEdit{In model-based settings, where the model parameters $P$ and $r$ are exactly known, the stationary distribution $\bm\pi^{\bm\mu}$ can be obtained by solving 
$\bm\pi^{\bm\mu} \bm{P}^{\bm\mu}=\bm\pi^{\bm\mu}$ and $\bm\pi^{\bm\mu}\boldsymbol{e}=1$, where $\boldsymbol{e}$ is a column vector of dimension $|\mathcal{S}|$ with all entries equal to one. Given $\bm{f}^{\bm{\mu}}$, $\bm{\pi}^{\bm\mu}$, and $\bm{P}^{\bm{\mu}}$, the value function $\bm{V}_{f}^{\bm{\mu}}$ can be calculated as \citep{xia2016generalized} 
\begin{equation}
    \label{eq:fund_matrix}
    \bm{V}_{f}^{\bm{\mu}}=(\boldsymbol{I}-\bm{P}^{\bm{\mu}}+\boldsymbol{e}\bm{\pi}^{\bm\mu})^{-1}\bm{f}^{\bm\mu},
\end{equation}
where $\boldsymbol{I}$ is an $|\mathcal{S}|$-dimensional identity matrix. The functions $Q_f^{\bm\mu}$ and $A_f^{\bm\mu}$ are then given by (\ref{eq:qf function-2 }) and (\ref{eq:af function}), respectively. When the model parameters are unknown, both the mean-variance performance function and the value function can be estimated from sample trajectories.}

\huEdit{We note that, although the performance and value functions can be calculated or estimated, policy optimization for MV-TSGs faces two key challenges. First, while the Poisson equation remains valid, the Bellman optimality equation does not hold for joint policies. This is because the reward function $f^{\bm\mu}$ is non-Markovian, incorporating the term $\eta^{\bm\mu}$, which depends on actions across both current and future stages. Second, the performance $J^{\bm{\mu}}$ is determined by the joint policy $\bm\mu$, and for each agent $i$, the policies of the other agents $\bm{\mu}_{-i}$ constitute part of its environment. When agents update their policies simultaneously, the resulting coupling effects make it difficult to ensure monotonic improvement of the joint policy. These challenges significantly limit the applicability of existing algorithms to MV-TSGs, highlighting the necessity of developing new approaches.}

\section{\huEdit{Policy Optimization with Known Models}}\label{sec:MAPI}
\huEdit{In this section, we first derive the mean-variance performance difference and derivative formulas using sensitivity-based optimization theory \citep{cao2007stochastic}. With the aid of these two key results, we propose the MV-MAPI algorithm by introducing the sequential update scheme, and establish its convergence. Finally, we analyze local geometric properties of first-order stationary points in MV-TSGs and propose a modified MV-MAPI method.}

\subsection{Analysis of Sensitive-based Optimization}
\huEdit{The framework of sensitivity-based optimization originates from perturbation analysis \citep{ho2012perturbation}. Its central idea is to exploit sensitivity information to guide policy improvement. Such information includes both performance derivatives and performance differences, enabling the optimization of general objectives in Markov models. Following this framework, we derive the difference and derivative formulas for MV-TSGs.
The proofs of Lemma~\ref{lemma:J-difference} and Lemma~\ref{lemma:J-derivative} are provided in Appendices~\ref{sec:appd_J-difference} and \ref{sec:appd_J-derivative}, respectively.}
\begin{lemma}[Performance Difference Formula for MV-TSGs]
    For any two joint policies $\bm{\mu}$, $\bm{\mu}' \in \mathcal{U}$, we have
    \begin{equation}
        J(\bm{\mu}') - J(\bm{\mu}) 
        = \mathbb{E}_{s \sim \pi^{\bm{\mu}'}, \bm{a} \sim \bm{\mu}'}[A_{f}^{\bm{\mu}}(s,\bm{a})] + \beta(\eta^{\bm\mu'} - \eta^{\bm\mu} )^2 .
        \label{eq:MVPDF-2}
    \end{equation}
    \label{lemma:J-difference}
\end{lemma}
\vspace{-2em}
\begin{lemma}[Performance Derivative Formula for MV-TSGs]
    Given any two joint policies $\bm{\mu},\bm{\mu}' \in \mathcal{U}$, we consider a mixed policy $\delta_{\bm\mu}^{\bm\mu'}$, 
    \begin{equation*}
        \delta_{\bm\mu}^{\bm\mu'}(\bm{a}|s) = (1-\delta)\bm{\mu}(\bm{a}|s)+\delta \bm{\mu}'(\bm{a}|s), \notag
    \end{equation*}
    where the joint action $\bm{a}$ follows $\bm{\mu}$ with probability $1-\delta$ and follows $\bm{\mu}'$ with probability $\delta$, $\delta \in [0,1]$. 
    Then,
    \begin{equation*}
        \frac{\mathrm{d}J{(\delta_{\bm\mu}^{\bm\mu'})}}{\mathrm{d}\delta}\Big|_{\delta=0} = \mathbb{E}_{s\sim \pi^{\bm{\mu}}, \bm{a}\sim \bm{\mu}'}[A_{f}^{\bm{\mu}}(s,\bm{a})]. \notag
    \end{equation*}
    \label{lemma:J-derivative}
\end{lemma}
\vspace{-2em}
In Lemma~\ref{lemma:J-difference}, we can observe that the second term on the r.h.s of (\ref{eq:MVPDF-2}) is always positive. This implies that if a joint policy $\bm{\mu}'$ is chosen such that the expected advantage function is non-negative at every state $s$, i.e., $\sum\limits_{\bm{a}} \bm\mu'(\bm{a}|s)[A_{f}^{\bm{\mu}}(s,\bm{a})] \ge 0$, the performance is guaranteed to improve. \huEdit{Lemma~\ref{lemma:J-derivative} describes the performance derivative at policy $\bm\mu$ towards another policy $\bm\mu'$, and it indicates the policy optimization direction in MV-TSGs. Together, these lemmas establish the analytical foundation for joint policy improvement, motivating the algorithmic developments in subsequent sections.}

\subsection{Optimization Method}
\huEdit{Lemma~\ref{lemma:J-difference} suggests a valid approach for updating the joint policy. However, in multi-agent settings, simultaneous policy updates are generally intractable due to environmental non-stationarity. To address this issue, we introduce the sequential update mechanism, in which agents update their policies one at a time according to a prescribed order, while the policies of all other agents remain fixed during each update.}

\huEdit{Before presenting the optimization method, we first define local NEs and establish the existence of deterministic (pure) optimal NEs in MV-TSGs, where $\mu_i: \mathcal{S} \mapsto \mathcal{A}_i$ for each agent $i$. These results indicate that solutions can be sought within the deterministic joint policy space $\mathcal{D}$, which enables the development of a policy iteration-type method. The proof of Theorem~\ref{threm:deterministic NE} is provided in Appendix~\ref{sec:appd_threm1}.}

\begin{definition}[Local Nash Equilibrium]
    \label{def:local_NE}
    In a stochastic game, a joint policy $\bm{\mu}^*=(\mu^*_1,\ldots,\mu^*_N) \in \mathcal{U}$ is a local Nash equilibrium if $\exists \bar{\delta}\in (0,1]$, for all $\delta \in (0,\bar{\delta}]$, we have
    \begin{equation*}
        J(\mu^*_i,\bm\mu^*_{-i}) \ge J(\delta_{\mu^*_i}^{\mu_i},\bm\mu^*_{-i}), \quad \forall \mu_i \in \mathcal{U}_i, ~ i \in \mathcal{N}, \notag
    \end{equation*}
     where $\delta_{\mu^*_i}^{\mu_i} = (1-\delta)\mu_i^* + \delta \mu_i$. The equilibrium is called a strict local NE if the inequality holds strictly. Moreover, when $\bar{\delta}=1$, the local NE becomes an NE. 
\end{definition} 

\begin{remark}
    We note that the policy of each agent $i$ is a mapping $\mu_i:\mathcal{S} \mapsto \Delta(\mathcal{A}_i)$, which can be represented as a vector of dimension $|\mathcal{S}||\mathcal{A}_i|$. Definition~\ref{def:local_NE} is motivated by the notion of a local optimum in $\mathbb{R}^n$ and is formulated in terms of the directional derivative. Specifically, given a function $f:F\subseteq\mathbb{R}^n\to \mathbb{R}$, where $F$ denotes the domain of $f$, a point $x^* \in F$ is a local maximum if there exists $\epsilon>0$ such that for all $ p \in (0,\epsilon]$ and any feasible direction $v$, it holds that $f(x^*)\ge f(x^*+pv)$. Furthermore, for a feasible direction $v$, the directional derivative is defined as $\nabla_v f(x)=\frac{\text{d}g(p)}{\text{d}p}|_{p=0}$, where $g(p)=f(x+pv)$.
     For more details,  see \citet[Chapter 6]{chong2013introduction}.
    
\end{remark}

\begin{theorem}
    MV-TSGs at least have a deterministic (pure) Nash policy, which achieves the maximum of the mean-variance performance function. 
    \label{threm:deterministic NE}
\end{theorem}

\huEdit{Next, we propose the MV-MAPI algorithm for MV-TSGs based on Lemma~\ref{lemma:J-difference} and the sequential update mechanism, as shown in Algorithm~\ref{alg:MVMAPI}. During each outer-loop iteration, a random permutation of all agents is generated. The agents then update their policies sequentially in this order. Following each update, the corresponding advantage function is recalculated.}

To analyze Algorithm~\ref{alg:MVMAPI}, we define the first-order stationary point in the form of a mixed joint policy for MV-TSGs. Similar definitions of stationary points are also introduced by \cite{leonardos2022global} and \cite{zhang2024gradient} for the directly parameterized Markov potential games. Subsequently, Theorem~\ref{threm:MAPI_convergence} demonstrates the convergence of Algorithm~\ref{alg:MVMAPI}.

\begin{algorithm}[H]
    \label{alg:MVMAPI}
    \caption{Mean-variance multi-agent policy iteration with monotonic improvement property}
    \SetAlgoLined
    \setstretch{0.9}

    Initialize a deterministic joint policy $\bm{\mu}^{(0)} = (\mu^{(0)}_1,\ldots,\mu^{(0)}_N)$.
    
    \For{$k=0,1,\ldots$}
    {

        Let $\hat{\bm{\mu}}^{(k,0)} =\bm{\mu}^{(k)} $.
        
        Draw a permutation $i_{1:N}$ of agents at random.
        
        \For{$h=1:N$}
        {
            For the policy $\hat{\bm{\mu}}^{(k,h-1)}$, compute the values of $\eta^{\hat{\bm{\mu}}^{(k,h-1)}}$,$f^{\hat{\bm{\mu}}^{(k,h-1)}}$,$J^{\hat{\bm{\mu}}^{(k,h-1)}}$,  respectively.
            
            Compute the values of $V_{f}^{\hat{\bm\mu}^{(k,h-1)}}(s)$ for all states $s$, $Q_{f}^{\hat{\bm\mu}^{(k,h-1)}}(s,\bm{a})$ and $A_{f}^{\hat{\bm\mu}^{(k,h-1)}}(s,\bm{a})$ for all state-action pairs $(s,\bm{a})$, respectively.
            
            Update the individual policy of agent $i_h$
            \begin{align}
                &\mu^{(k+1)}_{i_h}(s)=\mathop{\arg\max}\limits_{a_{i_h}} \big[A_{f}^{\hat{\bm{\mu}}^{(k,h-1)}}(s,a_{i_h},\bm{a}_{-i_h}) \big], \quad \bm{a}_{-i_h}\sim \hat{\bm{\mu}}^{(k,h-1)}_{-i_h},\forall s \in \mathcal{S}.
                \label{eq:max} \\
                &\text{(\small Let $\mu_{i_h}^{(k+1)}(s)=\mu_{i_h}^{(k)}(s)$ when $\mu_{i_h}^{(k)}(s) $ can already achieve max in (\ref{eq:max}).)} \notag
            \end{align}

            Update the joint policy $\hat{\bm{\mu}}^{(k,h)} =(\mu^{(k+1)}_{i_1}, \ldots,\mu^{(k+1)}_{i_h},\mu^{(k)}_{i_{h+1}},\ldots,\mu^{(k)}_{i_N})$.
        }
        \eIf{$\mu_i^{(k)} == \mu_i^{(k+1)}, \forall i \in \mathcal{N}$}
            {Done and break.}
            {$\bm{\mu}^{(k+1)} = \hat{\bm{\mu}}^{(k,N)}$.}
    }
    Return $\bm{\mu}^{(k)}$.
\end{algorithm}

\begin{definition}[First-order Stationary Point]
    A joint policy $\tilde{\bm\mu}=(\tilde{\mu}_1,\ldots,\tilde{\mu}_N)$ is a first-order stationary point in MV-TSGs if for any $\mu_i\in \mathcal{U}_i$ and $\delta_{\tilde{\mu}_i}^{\mu_i}=(1-\delta)\tilde{\mu}_i + \delta \mu_i$, we have
    \begin{equation}
    \frac{\mathrm{d}J{(\delta_{\tilde{\mu}_i}^{\mu_i},\tilde{\bm\mu}_{-i})}}{\mathrm{d}\delta}\Big|_{\delta=0} \le 0, \quad \forall i \in \mathcal{N}.
        \label{eq:stationay_neq}
    \end{equation} 
    \label{def:statationary policy}
\end{definition}

\begin{theorem}
    Algorithm~\ref{alg:MVMAPI} converges to a first-order stationary point monotonically.
    \label{threm:MAPI_convergence}
\end{theorem}
\proof{}
    \huEdit{With Lemma~\ref{lemma:J-difference}, in the sequential update process we have $J(\bm\mu^{(k)}) = J(\hat{\bm{\mu}}^{(k,0)}) \le J(\hat{\bm{\mu}}^{(k,1)}) \le \cdots \le J(\hat{\bm{\mu}}^{(k,N)}) = J(\bm\mu^{(k+1)})$, which demonstrates the monotonicity of Algorithm~\ref{alg:MVMAPI}.  Furthermore, as the reward function is bounded, so is the mean-variance performance function $J^{\bm\mu}$. Therefore, the monotonicity and convergence of Algorithm~\ref{alg:MVMAPI} are proved. }
    
    \huEdit{Assume that Algorithm~\ref{alg:MVMAPI} converges to $\bm\mu$. For any agent $i$, we have
    \begin{equation}
        \mathbb{E}_{\bm{a}_{-i}\sim \bm\mu_{-i}} [A_{f}^{\mu_i,\bm\mu_{-i}}(s,a_i,\bm{a}_{-i})] \le 0, \quad \forall s \in \mathcal{S}, a_i \in \mathcal{A}_i.
        \label{eq:local_br_le0}
    \end{equation}
     Considering the mixed policy $\delta_{\mu_i}^{\mu_i'}=(1-\delta)\mu_i +\delta \mu_i'$, $\mu_i' \in \mathcal{U}_i$, Lemma~\ref{lemma:J-derivative} and  (\ref{eq:local_br_le0}) jointly imply $\frac{\text{d}J(\delta_{\mu_i}^{\mu_i'},\bm\mu_{-i})} {\text{d}\delta}\Big|_{\delta=0} \le 0$. Then, we can conclude that Algorithm~\ref{alg:MVMAPI} converges to a first-order stationary point according to Definition~\ref{def:statationary policy}.}
\endproof
\begin{remark}
    All local NEs, including NEs and optimal joint policies, are first-order stationary points. However, the reverse is not true. More details are investigated in Section~\ref{sec:analysis of stationary points}.
\end{remark}

Although Algorithm~\ref{alg:MVMAPI} converges to a first-order stationary point, MV-MAPI is expected to converge rapidly, similar to policy iteration in traditional MDP theory. However, deriving a specific analysis of the algorithmic complexity of Algorithm~\ref{alg:MVMAPI} is difficult. \huEdit{This is because the algorithmic complexity of classical policy iteration in the average-reward setting remains an open problem \citep{ye2011simplex}.}

\subsection{Analysis of Stationary Points in MV-TSGs}\label{sec:analysis of stationary points}

\huEdit{In contrast to the known result that the first-order stationary points of standard discounted or average-reward TSGs coincide with NEs \citep{zhang2024gradient, cheng2024provable, lei2020asynchronous}, Theorem~\ref{theorem:stationary policy} demonstrates that the local geometry and landscape of stationary points in MV-TSGs are considerably more complex, and its proof is presented in Appendix~\ref{sec:appd_stationary policy}.}

\huEdit{The intuition is as follows. In single-agent MDPs, first-order stationary points coincide with the global optima under discounted or average-reward criteria \citep{bhandari2024global}. Under risk-sensitive criteria, however, the local property of the performance function is typically shaped by more intricate derivative structures. In the mean-variance setting considered here, first-order stationarity merely ensures that the first term on the r.h.s. of the difference formula in Lemma~\ref{lemma:J-difference} cannot be further improved along any policy direction, while the performance difference remains governed by the second term.}

\begin{theorem}
    For the first-order stationary point $\tilde{\bm\mu}$ in MV-TSGs, we have
    \begin{itemize}
        \item If $\tilde{\bm\mu}$ satisfies $\frac{\mathrm{d}J(\delta_{\tilde{\mu}_i}^{\mu_i},\tilde{\bm\mu}_{-i})}{\mathrm{d}\delta}\Big|_{\delta=0} < 0$ for any agent $i \in \mathcal{N}$ and $\delta_{\tilde{\mu}_i}^{\mu_i}=(1-\delta)\tilde{\mu}_i + \delta \mu_i$, $\mu_i \in \mathcal{U}_i$, then $\tilde{\bm\mu}$ is a strict local NE in MV-TSGs. 
        \item If there exist some agents $i$ and mixed policies $\delta_{\tilde{\mu}_i}^{\mu_i'}=(1-\delta)\tilde{\mu}_i + \delta \mu_i'$, $\mu_i'\in \mathcal{U}_i$, satisfy $\frac{\mathrm{d}J(\delta_{\tilde{\mu}_i}^{\mu_i'},\tilde{\bm\mu}_{-i})}{\mathrm{d}\delta}\Big|_{\delta=0} = 0$, a necessary and sufficient condition for the stationary point is a local NE is that: for these agents $i$ and $\mu'_i$,  $\exists \bar\delta \in (0,1], \forall \delta \in (0,\bar\delta]$, \textbf{the long-run average reward} holds that $\eta(\delta_{\tilde{\mu}_i}^{\mu_i'},\tilde{\bm\mu}_{-i})=\eta(\tilde{\mu}_i,\tilde{\bm\mu}_{-i})$. 
    \end{itemize}
    \label{theorem:stationary policy}
\end{theorem}

Theorem~\ref{theorem:stationary policy} implies that first-order stationary joint policies may serve as (strict) local NEs under certain conditions. Although the necessary and sufficient condition in the second term is intricate, it provides a better understanding of the geometry of the problem. For example, if there exist an agent $i$ and a policy $\mu'_i$ such that $\frac{\text{d}J(\delta_{\tilde{\mu}_i}^{\mu'_i},\tilde{\bm\mu}_{-i})}{\text{d}\delta}\Big|_{\delta=0} = 0$ and $\frac{\text{d}\eta(\delta_{\tilde{\mu}_i}^{\mu'_i},\tilde{\bm\mu}_{-i})}{\text{d}\delta}\Big|_{\delta=0} \neq 0$, the first-order stationary point $\tilde{\bm\mu}$ is not a local NE but an unstable saddle point. Furthermore, inspired by the second term in Theorem~\ref{theorem:stationary policy},  Corollary~\ref{coro:escape_saddle} shows that mean-variance performance can sometimes be further improved, even at saddle points.

\begin{corollary}
      For a first-order stationary point $\tilde{\bm\mu}$, if there exists some agent $i$ and policy $\mu'_i$, it holds that $\frac{\mathrm{d}J(\delta_{\tilde{\mu}_i}^{\mu'_i},\tilde{\bm\mu}_{-i})}{\mathrm{d}\delta}\Big|_{\delta=0} = 0$ and $ \eta(\tilde{\mu}_i, \tilde{\bm\mu}_{-i}) \neq \eta(\mu'_i, \tilde{\bm\mu}_{-i}) $, the joint policy can be improved by updating $\mu_i=\mu'_i$. 
    \label{coro:escape_saddle}
\end{corollary}

Corollary~\ref{coro:escape_saddle} follows directly from $\frac{\mathrm{d}J(\delta_{\tilde{\mu}_i}^{\mu'_i},\tilde{\bm\mu}_{-i})}{\mathrm{d}\delta}\Big|_{\delta=0} = 0$ and Lemma~\ref{lemma:J-difference}, based on the proof of Theorem~\ref{theorem:stationary policy}. Moreover, since Algorithm~\ref{alg:MVMAPI} searches policies over the finite set $\mathcal{D}$, the corollary indicates an approach to avoid entrapment at certain stationary points $\tilde{\bm\mu}$ and to obtain improved joint policies, as illustrated in Algorithm~\ref{alg:mod_MVMAPI}.

\begin{algorithm}[H]
    \label{alg:mod_MVMAPI}
    \caption{Modified mean-variance multi-agent policy iteration}
    \SetAlgoLined
    \setstretch{0.9}
    Run Algorithm~\ref{alg:MVMAPI} and obtain a converged joint policy $\tilde{\bm\mu}$.

    \For{$i=1,\ldots$,N}
    {
        Find a policy set $\mathcal{D}_i^{\tilde{\bm\mu}} = \left\{\mu_i:\forall s\in \mathcal{S},\mu_i(s)= \arg\max\limits_{a_i} \mathbb{E}_{\bm{a}_{-i} \sim \tilde{\bm{\mu}}_{-i}}[A_{f}^{\tilde{\bm\mu}}(s,a_i,\bm{a}_{-i})] \Big|\mu_i \neq \tilde{\mu}_i \right\}$.

        \eIf{$\mathcal{D}_i^{\tilde{\bm\mu}}$ is empty}
            {Continue.}
            {\For{each $\mu_i \in\mathcal{D}_i^{\tilde{\bm\mu}}$}
                {
                \If{$\eta(\mu_{i},\tilde{\bm\mu}_{-i}) \neq \eta(\tilde{\bm\mu})$}
                    {Go to step 1 with an initial joint policy $(\mu_{i},\tilde{\bm\mu}_{-i})$.}
                }
            }
    }

    Obtain $\tilde{\bm\mu}$ and $\{\mathcal{D}_i^{\tilde{\bm\mu}}\}_{i=1,\ldots,N}$.

\end{algorithm}

\huEdit{For a stationary point $\tilde{\bm\mu}$ from Algorithm~\ref{alg:MVMAPI}, Algorithm~\ref{alg:mod_MVMAPI} searches for deterministic policies $\mu_i \in \mathcal{D}_i$ other than $\tilde{\mu}_i$ that satisfies $\frac{\text{d}J{(\delta_{\tilde{\mu}_i}^{\mu_i},\tilde{\bm\mu}_{-i})}}{\text{d}\delta}\Big|_{\delta=0} = 0$, and gets the set $\mathcal{D}_i^{\tilde{\bm\mu}}$. If there exists $\mu_{i} \in \mathcal{D}_i^{\tilde{\bm\mu}} $ with $\eta(\mu_{i},\tilde{\bm\mu}_{-i}) \neq \eta(\tilde{\bm\mu})$, Corollary~\ref{coro:escape_saddle} ensures $\tilde{\bm\mu}$ can be improved by $(\mu_{i},\tilde{\bm\mu}_{-i})$. Consequently, Algorithm~\ref{alg:mod_MVMAPI} obtains an improved joint policy $\tilde{\bm\mu}$ and sets $\{\mathcal{D}_i^{\tilde{\bm\mu}}\}_{i=1,\ldots,N}$.}

\huEdit{We note that, for $\tilde{\bm\mu}$ obtained by Algorithm~\ref{alg:mod_MVMAPI}, the condition $\frac{\text{d}J{(\delta_{\tilde{\mu}_i}^{\mu_i},\tilde{\bm\mu}_{-i})}}{\text{d}\delta}\Big|_{\delta=0} = 0$ still holds for any policy $\mu_i \in \mathcal{D}_i^{\tilde{\bm\mu}}$, with $\eta(\tilde{\bm\mu})=\eta(\mu_i,\tilde{\bm\mu}_{-i})$, which implies that $J(\tilde{\bm\mu})=J(\mu_i,\tilde{\bm\mu}_{-i})$ by Lemma~\ref{lemma:J-difference}. Therefore, we further explore the local property of $\tilde{\bm\mu}$ within a policy space excluding $\left\{ (\mu_i,\tilde{\bm\mu}_{-i}),\mu_i \in \mathcal{D}_i^{\tilde{\bm\mu}}, \forall i \in \mathcal{N} \right\}$. To this end, we define a \emph{valid pruned joint policy space} $\widetilde{\mathcal{D}}$ and demonstrate the local properties of $\tilde{\bm\mu}$ by  Theorem~\ref{threm:pruned_space}.}


\begin{definition}
    A joint policy space $\widetilde{\mathcal{D}} \subset \mathcal{D}$ is a valid pruned joint policy space if an optimal joint policy of the MV-TSG can be obtained in $\widetilde{\mathcal{D}}$.
\end{definition}

\begin{theorem}
    For the joint policy $\tilde{\bm\mu}$ and sets $\{\mathcal{D}_i^{\tilde{\bm\mu}}\}_{i=1,\ldots,N}$ obtained by Algorithm~\ref{alg:mod_MVMAPI}, let $\mathcal{D}^{\tilde{\bm\mu}}:=\left\{ (\mu_i,\tilde{\bm\mu}_{-i}),\mu_i \in \mathcal{D}_i^{\tilde{\bm\mu}}, \forall i \in \mathcal{N} \right\}$. We have 
    \begin{itemize}
        \item $J({\tilde{\bm\mu}})=J({\bm\mu}), \forall \bm\mu \in \mathcal{D}^{\tilde{\bm\mu}}$. Moreover, the policy space defined by $\widetilde{\mathcal{D}}:=\mathcal{D}\setminus \mathcal{D}^{\tilde{\bm\mu}}$ is a valid pruned joint policy space. 
        \item The converged joint policy $\tilde{\bm\mu}$ is a strict local NE in the mixed joint policy space induced by $\widetilde{\mathcal{D}}$. 
    \end{itemize}
    \label{threm:pruned_space}
\end{theorem}
\proof{}
From the above analysis, we arrive at  $J(\tilde{\bm\mu})=J(\bm\mu)$ for all $ \bm\mu \in \mathcal{D}^{\tilde{\bm\mu}}$, which indicates that  no joint policy in $\mathcal{D}^{\tilde{\bm\mu}}$ outperforms $\tilde{\bm\mu}$. Then, $\widetilde{\mathcal{D}}$ is a valid pruned joint policy space.

Moreover, in the mixed policy space induced by $\widetilde{\mathcal{D}}$, $\frac{\text{d}J(\delta_{\tilde{\mu}_i}^{\mu_i},\tilde{\bm\mu}_{-i})}{\text{d}\delta}\Big|_{\delta=0} < 0$ holds for all $i \in \mathcal{N}$, where $\delta_{\tilde{\mu}_i}^{\mu_i}=(1-\delta)\tilde{\mu}_i + \delta \mu_i$ and  $(\mu_i,\tilde{\bm\mu}_{-i}) \in \widetilde{\mathcal{D}}$. Hence, by  Theorem~\ref{theorem:stationary policy}, $\tilde{\bm\mu}$ is a strict local NE in this mixed policy space. The proof is complete.
\endproof

\begin{remark}
\label{remark:2}
    \huEdit{We note that $\mathcal{D}^{\tilde{\bm\mu}}$ is typically empty, since any $\bm\mu \in \mathcal{D}^{\tilde{\bm\mu}}$ must satisfy (i) $\frac{\mathrm{d}J(\delta_{\tilde{\mu}_i}^{\mu_i},\tilde{\bm\mu}_{-i})}{\mathrm{d}\delta}\Big|_{\delta=0} = 0$ for some agent $i$ and (ii) $\eta^{\tilde{\bm\mu}}=\eta^{\bm\mu}$. When $\mathcal{D}^{\tilde{\bm\mu}}=\emptyset$, we have $\widetilde{\mathcal{D}}=\mathcal{D}$, and the joint policy $\tilde{\bm\mu}$ from Algorithm~\ref{alg:mod_MVMAPI} becomes a strict local NE in $\mathcal{U}$. This follows because, for any state $s$ and agent $i$, $\mathbb{E}_{\bm{a}_{-i}\sim \tilde{\bm\mu}_{-i}}A_f^{\tilde{\bm\mu}}(s,a_i,\bm{a}_{-i})<0$ when $a_i$ is not chosen by $\tilde{u}_i$, implying $\frac{\mathrm{d}J(\delta_{\tilde{\mu}_i}^{\mu_i},\tilde{\bm\mu}_{-i})}{\mathrm{d}\delta}\Big|_{\delta=0} = \mathbb{E}_{s\sim \pi^{\tilde{\bm\mu}}, a_i \sim \mu_i, \bm{a}_{-i}\sim \tilde{\bm\mu}_{-i}}[A_{f}^{\tilde{\bm\mu}}(s,a_i, \bm{a}_{-i})]<0$ for any $\mu_i \in \mathcal{U}_i$ other than $\tilde{\mu}_i$.}
\end{remark}

Additionally, we consider a specific scenario where $\eta$ is the same for all joint policies, as discussed in Corollary~\ref{coro:eq_eta_NE}. 

\begin{corollary}
    If the long-run average reward is the same for all joint policies $\bm\mu \in \mathcal{U}$ in the MV-TSG, that is, $\eta^{\bm\mu}$ is independent of $\bm\mu$, Algorithm~\ref{alg:MVMAPI} will converge to an NE.
    \label{coro:eq_eta_NE}
\end{corollary}

Corollary~\ref{coro:eq_eta_NE} holds because, when $\eta^{\bm\mu}$ is identical for all joint policies $\bm\mu$, the second term on the r.h.s of Equation~(\ref{eq:MVPDF-2}) vanishes. When Algorithm~\ref{alg:MVMAPI} converges to $\tilde{\bm\mu}$, each $\tilde{\mu}_i$ is a best response to $\tilde{\bm\mu}_{-i}$, implying that $\tilde{\bm\mu}$ is an NE. One example satisfying this condition is presented in Section~\ref{sec:exp_results}.

\begin{figure}[htbp!]
    \centering
    \includegraphics[width=0.80\linewidth]{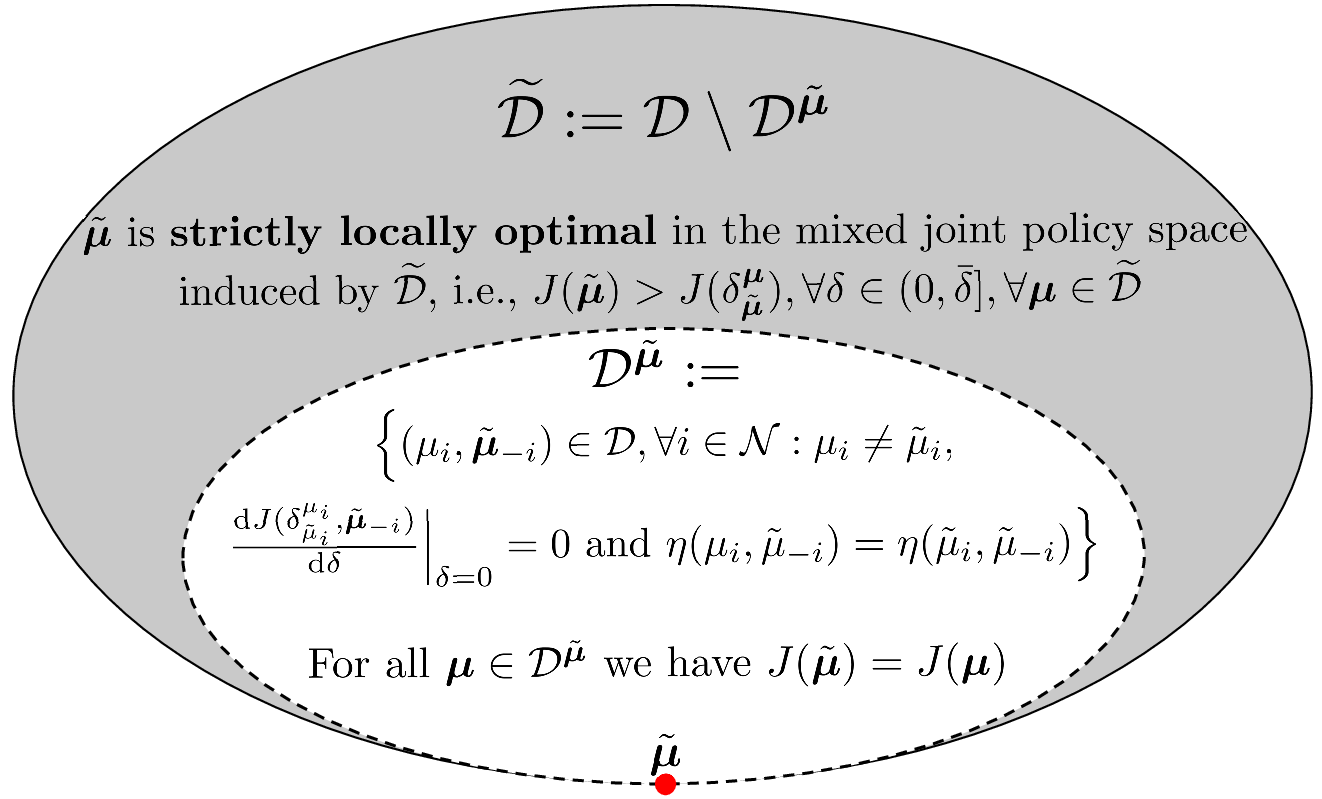}
    \caption{The optimality of the joint policy obtained by Algorithm~\ref{alg:mod_MVMAPI}.}
    \label{fig:divide_space}
\end{figure}

\huEdit{The above results suggest that converged stationary points are typically (strict) local NEs. However, their qualities are not specified. Theorem~\ref{theorem:strict local optima} characterizes the quality of strict local NEs and its proof is presented in Appendix~\ref{sec:app_ne_localoptimal}. Together with Theorem~\ref{threm:pruned_space} and Remark~\ref{remark:2}, we further provide Corollary~\ref{corollary:local optimal} and Figure~\ref{fig:divide_space} to demonstrate the local optimality of the joint policy obtained by Algorithm~\ref{alg:mod_MVMAPI}.}
\begin{theorem}
    In MV-TSGs, a strict local NE $\bm\mu^*$ is equivalent to a \textbf{strict local maximum} of the mean-variance performance function, i.e., $\exists \bar\delta \in (0,1] , \forall \delta \in (0,\bar\delta]$ we have $J(\bm\mu^*) > J(\delta_{\bm\mu^*}^{\bm\mu})$,$\forall \bm\mu\in \mathcal{U}$.
    \label{theorem:strict local optima}
\end{theorem}

\begin{corollary}
\label{corollary:local optimal}
    The joint policy $\tilde{\bm\mu} $ obtained by Algorithm~\ref{alg:mod_MVMAPI} is usually a strict local optimum in $\mathcal{U}$, and it is guaranteed to be a strict local optimum in the mixed joint policy space induced by the valid pruned joint policy space $\widetilde{\mathcal{D}}=\mathcal{D}\setminus \mathcal{D}^{\tilde{\bm\mu}}$; that is, there exists $ \bar{\delta}\in (0,1]$ such that for all  $\delta \in (0,\bar{\delta}]$,  $J(\tilde{\bm\mu})>J(\delta_{\tilde{\bm\mu}}^{\bm\mu})$ for any $ \bm\mu \in \widetilde{\mathcal{D}}$.
\end{corollary}

\section{Mean-Variance Multi-Agent Reinforcement Learning}\label{sec:MV-MARL}
\huEdit{In contrast to Algorithm~\ref{alg:MVMAPI}, which is limited to small-scale MV-TSGs with known environmental models, this section proposes a mean-variance MARL algorithm to tackle MV-TSGs with unknown environmental parameters, such as transition and reward functions. Our approach follows the algorithmic paradigm introduced by \cite{zhong2024heterogeneous}. By parameterizing the policy and value functions with neural networks, the proposed method scales efficiently to larger state spaces. It is worth noting that the MV-TSGs studied in this work differ substantially from the standard discounted setting studied in \citet{zhong2024heterogeneous}.}

\subsection{Sequential Update Scheme in Centralized Training}

\huEdit{In Algorithm~\ref{alg:MVMAPI}, each inner-loop iteration involves $N$ policy evaluations. To improve sample efficiency, our algorithm follows the CTDE framework, and updates policies of all agents merely rely on trajectory data collected under a common joint policy $\bm\mu$. During centralized training, agents update their policies sequentially, with each agent able to access and consider all preceding agents’ updates.}

\huEdit{To describe the sequential update in our MARL algorithm, we first introduce the following definitions. Note that these definitions are associated with the actions of a set of agents and a common joint policy $\bm\mu$, which differ from the variables specified in Algorithm~\ref{alg:MVMAPI}.}

\begin{definition}
    For an ordered subset of agents $i_{1:h}$ and the complement subset $-i_{1:h}$, we define the \textbf{multi-agent state-action value function for MV-TSGs} as
    \begin{equation*}
        Q_{f,i_{1:h}}^{\bm{\mu}}(s,\bm{a}_{i_{1:h}}):=\mathbb{E}_{\bm{a}_{-i_{1:h}}\sim \bm{\mu}_{-i_{1:h}}}\left[Q_{f}^{\bm{\mu}}(s,\bm{a}_{i_{1:h}},\bm{a}_{-i_{1:h}})\right]. \notag
    \end{equation*} 
    Furthermore, the \textbf{multi-agent advantage function for agent $i_h$} is
    \begin{equation*}
        A_{f,i_{h}}^{\bm{\mu}}(s,\bm{a}_{i_{1:h-1}},a_{i_{h}}):= Q_{f,i_{1:h}}^{\bm{\mu}}(s,\bm{a}_{i_{1:h-1}},a_{i_{h}})-Q_{f,i_{1:h-1}}^{\bm{\mu}}(s,\bm{a}_{i_{1:h-1}}), \notag
    \end{equation*}
    where $Q_{f,i_{1:0}}^{{\bm{\mu}}}(s,\bm{a}_{i_{1:0}}) = V_{f}^{\bm{\mu}}(s)$.
    \label{def:ma-QA function}
\end{definition}
\huEdit{The state-action value function $Q_{f,i_{1:h}}^{\bm{\mu}}(s,\bm{a}_{i_{1:h}})$ evaluates the value of agents $i_{1:h}$ taking actions $\bm{a}_{i_{1:h}}$ in state $s$ while other agents follow the joint policy $\bm\mu_{-i_{1:h}}$. The advantage function $A_{f,i_{h}}^{\bm{\mu}}(s,\bm{a}_{i_{1:h-1}},a_{i_{h}})$ evaluates the advantage of agent $i_{h}$ taking action $a_{i_{h}}$ in state $s$ given that agents $i_{1:h-1}$ have taken actions $\bm{a}_{i_{1:h-1}}$ and the rest of agents follow the joint policy $\bm\mu_{- i_{1:h}}$.}

\huEdit{Subsequently, the accumulated multi-agent advantage $A_{f,i_{1:h}}^{\bm\mu}(s,\bm{a}_{i_{1:h}})$ is given by 
\begin{equation}
\label{eq:ad_decompostion}
A_{f,i_{1:h}}^{\bm\mu}\big(s,\bm{a}_{i_{1:h}}\big)=\sum_{j=1}^h A_{f,i_j}^{\bm\mu}\big(s,\bm{a}_{i_{1:j-1}},a_{i_j}\big),
\end{equation}
which represents the advantage of agents $i_{1:h}$ taking action $\bm{a}_{i_{1:h}}$ in state $s$ relative to the value function $V_f^{\bm\mu}(s)$. Then, given a permutation $i_{1:N}$ of all agents,
the first term of the r.h.s in (\ref{eq:MVPDF-2}) can be rewritten as
\begin{equation}
\label{eq:MVPDF-firstterm}
    \mathbb{E}_{s \sim \pi^{\bm{\mu}'}, {\bm{a}} \sim \bm{\mu}'}[A_{f}^{\bm{\mu}}(s,{\bm{a}})]=\mathbb{E}_{s \sim \pi^{\bm{\mu}'}, \bm{a}_{i_{1:N}}\sim \bm\mu'}\Big[\sum_{h=1}^N A_{f,i_h}^{\bm\mu}\big(s,\bm{a}_{i_{1:h-1}},a_{i_h}\big)\Big].
\end{equation}
We can sequentially select each agent action $a'_{i_h}$ such that $A_{f,i_h}^{\bm\mu}(s,\bm{a}'_{i_{1:h-1}}, a'_{i_h}) \ge 0$, given that $\mathbb{E}_{a_{i_h} \sim \mu_{i_h}}\Big[A_{f,i_h}^{\bm\mu}(s,\bm{a}'_{i_{1:h-1}}, a_{i_h})\Big] = 0$. Moreover, any $A_{f,i_h}^{\bm\mu}(s,\bm{a}'_{i_{1:h-1}}, a'_{i_h}) > 0$ ensures that $A_{f,i_{1:N}}^{\bm\mu}(s, \bm{a}'_{i_{1:N}})>0$, thereby improving the MV-TSG performance.
Consequently, during training, the policy of agent $i_h$ can be updated by 
\begin{equation}
    \label{eq:max_A}\mathop{\arg\max}_{a_{i_h}\in \mathcal{A}_{i_h}}A_{f,i_h}^{\bm\mu}(s,\bm{a}'_{i_{1:h-1}}, a_{i_h}), \quad \forall s \in \mathcal{S}.
\end{equation}
}

The remaining key step is to estimate the optimization objective $A_{f,i_h}^{\bm\mu}$ for each agent $i_h$ using trajectory data collected under the joint policy $\bm{\mu}$. The following proposition formalizes this result, and its proof is provided in  Appendix~\ref{sec:apped_prop2}.

\begin{proposition}
    Let $\bm\mu = (\mu_1,\ldots,\mu_N)$ be a joint policy, and $A_{f}^{\bm\mu}(s,\bm{a})$ be its advantage function. For a given order set $i_{1:N}$, let $\bm\mu'_{i_{1:h-1}} = (\mu'_{i_1},\ldots, \mu'_{i_{h-1}})$ be some other joint policy of agents $i_{1:h-1}$, and $\hat{\mu}_{i_h}$ be some other policy of agent $i_h$. Then, for every state $s$,
    \begin{align}
        \mathbb{E}_{\bm{a}_{i_{1:h-1}} \sim \bm\mu'_{i_{1:h-1}}, a_{i_h} \sim \hat{\mu}_{i_h}} &[A_{f,i_h}^{\bm\mu}(s, \bm{a}_{i_{1:h-1}},a_{i_h})] \notag\\
        &= \mathbb{E}_{\bm{a}\sim \bm\mu} \Big[
        \big(\frac{\hat{\mu}_{i_h} (a_{i_h}|s)}{\mu_{i_h} (a_{i_h}|s )}-1\big)  \frac{\bm\mu'_{i_{1:h-1}}(\bm{a}_{i_{1:h-1}}|s)}{\boldsymbol{\mu}_{i_{1:h-1}}(\bm{a}_{i_{1:h-1}}|s)} A_{f}^{\bm\mu}(s,\bm{a})  \Big]. \notag
    \end{align}
    \label{prop:A_estimator}
\end{proposition}
Proposition~\ref{prop:A_estimator} indicates that  given an advantage function estimator $\hat{A}_{f}^{\bm\mu}(s,\bm{a})$, we can estimate $\mathbb{E}_{\bm{a}_{i_{1:h-1}} \sim \bm\mu'_{i_{1:h-1}}, a_{i_h} \sim \hat{\mu}_{i_h}} [A_{f,i_h}^{\bm\mu}(s, \bm{a}_{i_{1:h-1}},a_{i_h})]$ with an estimator of 
\begin{equation*}
  \mathbb{E}_{{\bm{a}}\sim \bm\mu}\Big[\big(\frac{\hat{\mu}_{i_h}\left(a_{i_h}|s \right)}{\mu_{i_h}\left(a_{i_h}|s \right)}-1 \big)M_{f,i_{1:h}}^{\bm\mu}(s,\bm{a})\Big],  \notag
\end{equation*}
where $ M_{f,i_{1:h}}^{\bm\mu}(s,\bm{a})=\frac{\bm\mu'_{i_{1:h-1}}(\bm{a}_{i_{1:h-1}}|s)}{\bm{\mu}_{i_{1:h-1}}(\bm{a}_{i_{1:h-1}}|s)} \hat{A}_{f}^{\bm\mu}(s,\bm{a})$ and $\bm\mu'_{i_{1:0}}=\bm\mu$.
These definitions and results lay the foundation for centralized training with the sequential update. 

\subsection{Multi-Agent Trust Region Policy Optimization for MV-TSGs}
In the approximate setting, due to the estimation and approximation error, it is unavoidable that given a joint policy $\bm\mu$, the advantage function is negative for some agent $i_h$ and state $s$, i.e., $A_{f,i_h}^{\bm\mu}(s,\bm{a}'_{i_{1:h-1}}, a_{i_h}) < 0, \forall a_{i_h}\in\mathcal{A}_{i_h}$. Moreover, the term $\pi^{\bm\mu'}$ in (\ref{eq:MVPDF-firstterm}) depends on the next joint policy $\bm\mu'$. These observations make it difficult to optimize policies directly by (\ref{eq:max_A}). 

To address these issues, we extend the idea of trust region methods \citep{schulman2015trust,zhang2021policy} to the centralized training phase and propose the MV-MATRPO algorithm. For the policy update of each agent, our goal is to maximize a surrogate objective within a local trust region. 

\huEdit{Before introducing the surrogate objective for each agent and presenting the algorithm, we first present the average-reward trust region lemma \citep{zhang2021policy}. Subsequently, we present a performance improvement lower bound for MV-TSGs in Theorem~\ref{threm:mv_bounds}, and its proof is presented in Appendix~\ref{sec:app_joint_bound}.}

\begin{lemma}[Theorem~1, \cite{zhang2021policy}]
    \label{lemma:trpo_joint}
        Let $\bm\mu$ be the current joint policy and $\bm\mu'$ be any other joint policy. We define $\mathcal{L}^{\bm\mu}(\bm\mu') = \mathbb{E}_{s \sim \pi^{\bm\mu}, \bm{a} \sim {\bm\mu'}} [A^{\bm\mu}(s,\bm{a})]$, where $A^{\bm\mu}$ is the average-reward advantage function.
        With the ergodicity assumption, the following bounds hold:
    \begin{equation*} 
        \eta(\bm\mu')-\eta(\bm\mu) \ge \mathcal{L}^{\bm\mu}(\bm\mu') - 2(\kappa^*-1)\epsilon_{\eta}D_{\emph{TV}}(\bm\mu',\bm\mu), 
    \end{equation*}
    where $\epsilon_\eta=\max\limits_s \big|\mathbb{E}_{\bm{a}\sim \bm\mu'}[A^{\bm\mu}(s,\bm{a})] \big|$,  $\kappa^*=\max\limits_{\bm\mu}\kappa^{\bm\mu}$, $\kappa^{\bm\mu}$ is the Kemeny's constant \citep{kemeny1969finite} for a given $\bm\mu$, $D_{\emph{TV}}(\bm\mu',\bm\mu)= \mathbb{E}_{s \sim \pi^{\bm\mu}} D_{\emph{TV}}(\bm\mu'(\cdot | s) \| \bm\mu(\cdot |s))$ is the total variation divergence, which is a measure of distance between two distributions.
    \end{lemma}

\begin{theorem}
    Let $\bm\mu$ be the current joint policy and $\bm\mu'$ be any other joint policy. We define the surrogate objective function $\mathcal{L}_{f}^{\bm\mu}(\bm\mu') = \mathbb{E}_{s \sim \pi^{\bm\mu}, \bm{a} \sim {\bm\mu'}} [A_{f}^{\bm\mu}(s,\bm{a})]$. The following bounds hold in MV-TSGs:
    \begin{equation}
        J(\bm\mu') - J(\bm\mu) 
        \ge \mathcal{L}_{f}^{\bm\mu}(\bm\mu') - 2(\kappa^* - 1) \epsilon_f D_{\emph{TV}}(\bm\mu',\bm\mu) + \beta H^2, 
        \label{eq:mv_bound}
    \end{equation}    
    where $\epsilon_{f}=\max\limits_s \big|\mathbb{E}_{\bm{a} \sim \bm\mu'}[A_{f}^{\bm\mu}(s,\bm{a})] \big|$, $H =\max(0, \mathcal{L}^{\bm\mu}(\bm\mu')-2(\kappa^*-1)\epsilon_{\eta}D_{\emph{TV}}(\bm\mu',\bm\mu), -\mathcal{L}^{\bm\mu}(\bm\mu')- 2(\kappa^*-1)\epsilon_{\eta}D_{\emph{TV}}(\bm\mu',\bm\mu))$, $\mathcal{L}^{\bm\mu}(\bm\mu'), \epsilon_{\eta}, \kappa^*$ as defined in Lemma~\ref{lemma:trpo_joint}.
    \label{threm:mv_bounds}    
\end{theorem}
Theorem~\ref{threm:mv_bounds} suggests that as the trust region tightens, i.e., $\epsilon_f \to 0$, the sign of the performance difference can be determined by the first-order term $\mathcal{L}_{f}^{\bm\mu}(\bm\mu')$. This provides the theoretical foundation for applying trust region methods to long-run mean-variance optimization problems.

Next, we want to express the r.h.s of (\ref{eq:mv_bound}) as the sum of individual objectives for all agents, and we define the following surrogate objective for each agent.

\begin{definition}
    Let $\bm\mu$ be a joint policy, $\bm\mu'_{i_{1:h-1}}= (\mu'_{i_1},\ldots,\mu'_{i_{h-1}})$ be some other joint policy of agents $i_{1:h-1}$, and $\hat{\mu}_{i_h}$ be some other policy of agent $i_h$. Then 
    \begin{equation*}
        \mathcal{L}_{f,i_{1:h}}^{\bm\mu}(\bm\mu'_{i_{1:h-1}}, \hat{\mu}_{i_h}) = \mathbb{E}_{s \sim \pi^{\bm\mu}, \bm{a}_{i_{1:h-1}}\sim \bm\mu'_{i_{1:h-1}}, a_{i_h} \sim \hat{\mu}_{i_h}} [A_{f,i_h}^{\bm\mu}(s, \bm{a}_{i_{1:h-1}}, a_{i_h})]. 
    \end{equation*}
    \label{def:ma_surrogate}
\end{definition}

Combining Equation~(\ref{eq:ad_decompostion}), Theorem~\ref{threm:mv_bounds}, and Definition~\ref{def:ma_surrogate}, we replace the TV divergence with KL divergence as most trust region methods do, and derive the bound for the joint policy update as Theorem~\ref{threm:ma_bounds} demonstrates. The proof is presented in Appendix~\ref{sec:app_ma_bound}.

\begin{theorem}
    Let $\bm\mu$ be a joint policy. Then, for any joint policy $\bm\mu'$, we have
    \begin{equation*}
        J(\bm\mu')\ge J(\bm\mu) + \sum\limits_{h=1}^N [\mathcal{L}_{f,i_{1:h}}^{\bm\mu}(\bm\mu'_{i_{1:h-1}}, \mu'_{i_h}) - W(\mu'_{i_h}, \mu_{i_h})]. \notag
    \end{equation*}
    where $W(\mu'_{i_h}, \mu_{i_h}) = (\kappa^*-1)\epsilon_f  \sqrt{2\mathbb{E}_{s\sim \pi^{\bm\mu}} D_{\emph{\text{KL}}}(\mu'_{i_h}(\cdot|s)\|\mu_{i_h}(\cdot|s)) }$.
    \label{threm:ma_bounds}
\end{theorem}
Theorem~\ref{threm:ma_bounds} provides theoretical guarantees for the multi-agent trust region method based on the sequential update mechanism for MV-TSGs. Namely, if agents sequentially update their policies by
\begin{equation}
    \mathop{\arg\max}_{\mu'_{i_h}} [\mathcal{L}_{f,i_{1:h}}^{\bm\mu}(\bm\mu'_{i_{1:h-1}}, \mu'_{i_h}) - W(\mu'_{i_h}, \mu_{i_h})],
    \label{eq:mv_improvement}
\end{equation}
the joint policy is guaranteed to be improved, or at least remain non-decreasing by retaining the
current policy, i.e.,  $\mu'_{i_h} = \mu_{i_h}$.

\subsection{Implementation with Neural Network Parameterization}\label{sec:implementation details} 
To implement the policy update process in practical settings, for the ($k+1$)-th iteration of the sequential update, the policy $\mu_i^{(k)}$ of each agent $i$ and the value function $V_{f}^{\bm\mu^{(k)}}$ are parameterized by neural networks with parameters $\theta^{(k)}_i$ (actor networks) and $\phi^{(k)}$ (critic networks), respectively.  To ensure the algorithm is computationally tractable, the unconstrained optimization problem (\ref{eq:mv_improvement}) is reformulated as a constrained problem. Let  $\mu^{\theta_{i}^{(k)}}$ denote $\mu_i^{(k)}$, $V_f^{\phi^{(k)}}$ denote $V_{f}^{\bm\mu^{(k)}}$ and $\theta^{(k)}=(\theta_1^{(k)},\ldots,\theta_N^{(k)})$. With Theorem~\ref{threm:ma_bounds},  given a permutation of agents $i_{1:N}$, agent $i_{h \in \{1,\ldots,N \} }$ can sequentially optimize its policy parameter $\theta_{i_h}^{(k+1)}$ by maximizing the constrained objective:
\begin{align}
    \theta_{i_h}^{(k+1)} = \mathop{\arg\max}\limits_{\theta_{i_h}} \mathbb{E}_{s \sim \pi^{{\theta^{(k)}}}, \bm{a}_{i_{1:h-1}}\sim \bm\mu^{\theta_{i_{1:h-1}}^{(k+1)}}, a_{i_h} \sim \mu^{\theta_{i_h}}} [A_{f,i_h}^{{\theta^{(k)}}}(s,\bm{a}_{1:h-1}, a_{i_h})], \notag \\
    \text{s.t.} \quad \mathbb{E}_{s \sim \pi^{{\theta^{(k)}}}}[D_\text{KL}(\mu^{\theta^{(k)}_{i_h}}(\cdot|s), \mu^{\theta_{i_h}}(\cdot|s))] \le \epsilon,
    \label{eq:trust_indiv}
\end{align}
where $\epsilon$ is the threshold hyperparameter.

We use the method proposed by \cite{zhong2024heterogeneous} to solve the above optimization problem. More details are provided in Appendix~\ref{sec:appd_implementation details}.
With Proposition~\ref{prop:A_estimator}, the objective functions for each agent can be estimated with no bias based on the advantage function $\hat{A}_{f}^{{\theta^{(k)}}}(s,\bm{a})$,
which can be estimated using the generalized advantage estimation \citep{schulman2016high}. Typically, we have
\begin{equation}
    \hat{A}_{f}^{{\theta^{(k)}}}(s_n,\bm{a}_n)=\sum\limits_{t=n}^{T-1}\lambda_{t-n}(\hat{f}^{\theta^{(k)}}(s_t,\bm{a}_t) - \hat{J}^{{{\theta^{(k)}}}}+V_f^{\phi^{(k)}}(s_t) - V_f^{\phi^{(k)}}(s_{t+1})),
    \label{eq:compute_advantage}
\end{equation}
\huEdit{where $\lambda$ is the hyper-parameter to trade-off bias and variance, $T$ is the length of trajectories.}

Since we consider the long-run average performance in this paper, we adopt the average value constraint (AVC) proposed by \cite{maaverage} to assist in estimating the target value function $\hat{V}_f^{\phi^{(k)}}$ and stabilizing the value learning. The value function network is updated using the loss function
\begin{equation*}
     \frac{1}{BT}\sum\limits_{b=1}^B\sum\limits_{t=0}^T \big( V_f^{\phi}(s_t) - \hat{V}_f^{\phi^{(k)}}(s_t)  \big)^2, \notag
\end{equation*}
where $B$ is the number of trajectories collected. The detailed pseudo code of MV-MATRPO is presented in Algorithm~\ref{alg:MVTRPO}.

Moreover, in scenarios with larger state and action spaces, we can follow the route of Proximal Policy Optimization (PPO) \citep{schulman2017proximal} to mitigate the computational burden. The problem listed in  (\ref{eq:trust_indiv}) can be solved by focusing solely on the first-order derivative and sequentially optimizing the policy parameter $\theta^{(k+1)}_{i_h}$ through maximizing the clipping objective of 
\begin{equation}
\mathbb{E}_{s\sim \pi^{{\theta^{(k)}}},\bm{a} \sim \bm{\mu}^{\theta^{(k)}}}\left[\min
\left(\frac{\mu^{\theta_{i_h}}\left(a_{i_h}|s\right)}{\mu^{\theta^{(k)}_{i_h}}\left(a_{i_h}|s\right)}M_{f,i_{1:h}}^{{\theta^{(k)}}}\left(s,\bm{a}\right),\text{clip}\left(\frac{\mu^{\theta_{i_h}}\left(a_{i_h}|s\right)}{\mu^{\theta^{(k)}_{i_h}}\left(a_{i_h}|s\right)},1\pm\epsilon\right)M_{f,i_{1:h}}^{{\theta^{(k)}}}\left(s,\bm{a}\right)\right)\right], \notag
\end{equation}
where the \textit{clip} function clips the first argument by the lower and upper bounds denoted by the second and third arguments, respectively.

\begin{algorithm}[H]
    \label{alg:MVTRPO}
    \caption{Mean-variance multi-agent trust region policy optimization with parameterization}
    \SetAlgoLined
    \setstretch{0.9}
    
    \textbf{Input:} Stepsize $\alpha$, number of agents $N$, iterations $K$, episode length $T$.

    \textbf{Initialize:} Policy function (actor) networks $\{\theta^{(0)}_i, \forall i \in \mathcal{N} \}$, value function (critic) network $\{\phi^{(0)} \}$, replay buffer $\mathcal{B}$. Set $\hat{\eta}^{}=0, \hat{\zeta}=0, \hat{J}=0$.
    
    \For{$k=0,\ldots,K-1$}
    {

        Collect a set of trajectories by running the joint policy $\bm\mu^{\theta^{(k)}} = (\mu^{\theta^{(k)}_1}, \ldots, \mu^{\theta^{(k)}_N})$.

        Push transitions $\{(s_t,\{a_{i,t} \}_{i=0}^N,s_{t+1},r_t), t \in \{0,1,\ldots,T-1\}$ into $\mathcal{B}$.

        Update $\hat{\eta} \gets (1-\alpha)\hat{\eta}+\alpha\frac{1}{BT} \sum\limits_{b=1}^B \sum\limits_{t=0}^{T-1}r_t $.

        Update $\hat{\zeta} \gets (1-\alpha)\hat{\zeta}+\alpha\frac{1}{BT} \sum\limits_{b=1}^B \sum\limits_{t=0}^{T-1}(r_t-\hat{\eta})^2 $. 

        Compute the average mean-variance performance function $\hat{J}$.   
        
        Compute $\hat{f}(s_t, \bm{a}_t)$ and $\hat{A}_f^{{\theta^{(k)}}}(s_t, \bm{a}_t)$, respectively, at all time steps.
        
        Compute $\hat{V}_f^{\phi^{(k)}}(s_t)$ for all time steps using AVC.

        Draw a random permutation of agents $i_{1:N}$.

        Set $M_{f,i_1}^{{\theta^{(k)}}}(s,\bm{a}) = \hat{A}_{f}^{{\theta^{(k)}}}(s,\bm{a}), \forall s \in \mathcal{S}, \forall \bm{a}\in \mathcal{A}$.

        \For{agent $i_h=i_1,\ldots,i_N$}
        {
        Solve the optimization problem (\ref{eq:trust_indiv}) and get the policy $\mu^{\theta^{(k+1)}_{i_h}}$.    
    
            \If{$i_h \neq i_N$}
            {Compute $M_{f,i_{1:h+1}}^{{\theta^{(k)}}}(s,\bm{a})= \frac{\mu^{\theta^{(k+1)}_{i_h}}\left(a_{i_h}|s\right)}{\mu^{\theta^{(k)}_{i_h}}\left(a_{i_h}|s\right)} M_{f,i_{1:h}}^{{\theta^{(k)}}}(s,\bm{a}), \forall s \in \mathcal{S}, \forall \bm{a}\in \mathcal{A} $.
            }
        }
        Update the critic network by following the formula:
        \begin{equation*}
            \phi_{k+1} = \mathop{\arg\min}_{\phi} \frac{1}{BT}\sum\limits_{b=1}^B\sum\limits_{t=0}^T \big( V_f^{\phi}(s_t) - \hat{V}_f^{\phi^{(k)}}(s_t)  \big)^2.
        \end{equation*}       
    }

\end{algorithm}

\section{Applications in Energy Management}\label{sec:exp_results}
In this section, we use the energy management problem of MMSs as an example to demonstrate the effectiveness of our algorithms. We have significantly simplified the parameter settings and engineering constraints, ensuring the key idea of this example is concise and easy to follow.

\begin{figure}[H]
    \centering
    \includegraphics[width=12cm]{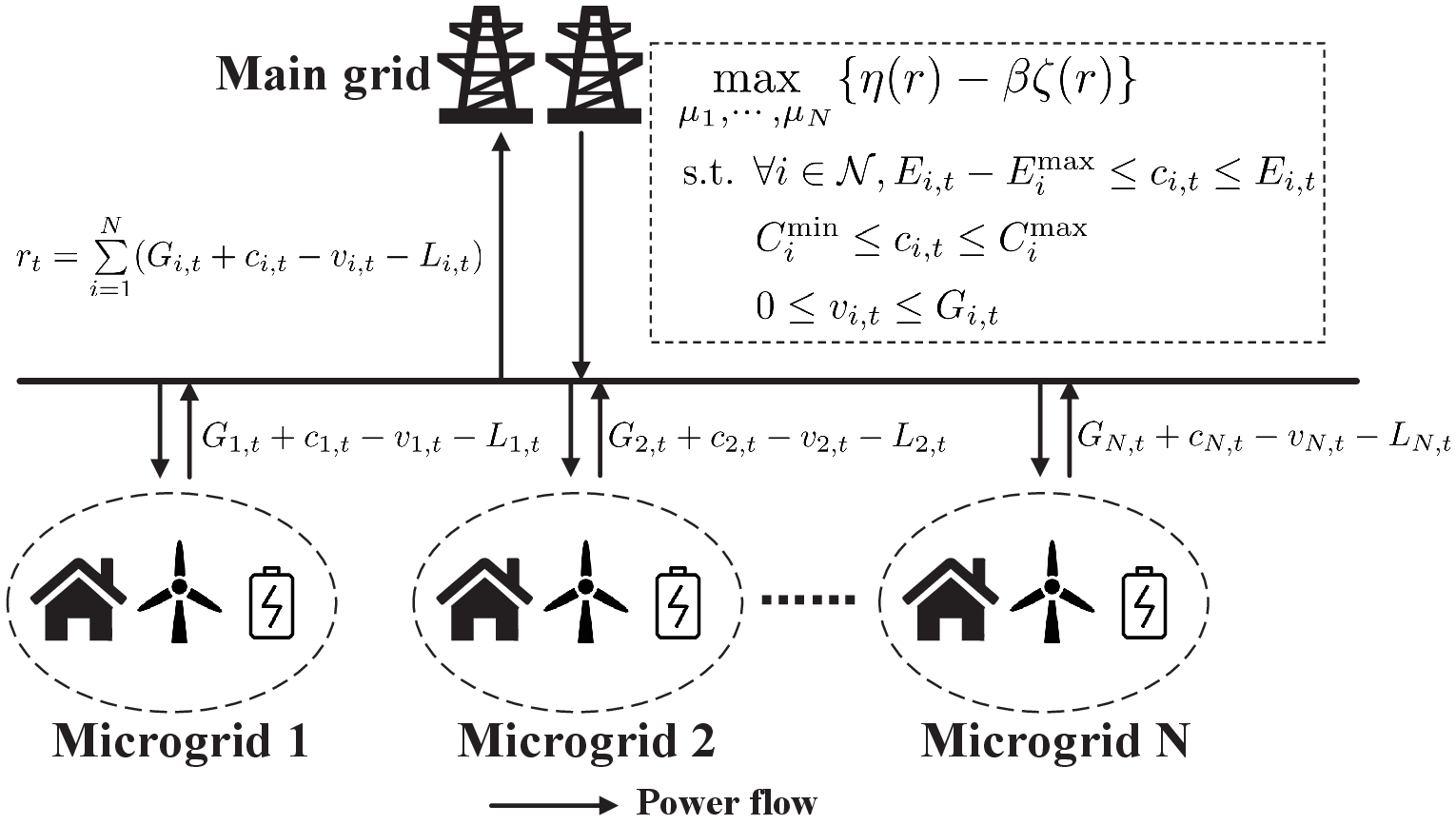}
    \caption{Architecture of a grid-connected multiple microgrids system.}\label{fig:networked microgrids}
    \end{figure}
    \vspace{-3mm}

\huEdit{We consider a grid-connected MMS comprising $N$ microgrids, illustrated in Figure~\ref{fig:networked microgrids}. Each microgrid may be equipped with a renewable energy generator, a demand load unit, and a controllable storage unit, which indicates microgrids
can both generate and consume power. We assume that abandoning generated power is allowed if necessary, and then each microgrid has an energy management policy $\mu_i$ for controlling renewable energy generators and storage units.}

\huEdit{The microgrids coordinate to regulate the exchanged power between the MMS and the main grid. Positive exchanged power means that the MMS sells the power to the main grid and gets profits, while negative exchanged power indicates that it buys power and incurs costs. The volatility of exchanged power indicates power supply stability, which can be characterized by the long-run variance. Then, the objective of the MMS is to establish optimal energy management policies for each microgrid to maximize the mean-variance of the exchanged power. The details of variable definitions and parameter settings are presented in Appendix~\ref{sec:app_setting_details}.}

\huEdit{ We conduct two sets of experiments to evaluate our algorithms. The first tests MV-MAPI in a small-scale, model-based scenario, analyzing the effects of update orders and initial policies. The second applies MV-MATRPO to two model-free scenarios: one replicating the first setting and another of significantly larger scale. All experiments were conducted on a machine equipped with an AMD 3995WX CPU, 256 GB of memory, and an Nvidia GeForce GTX 4090 GPU.}

\subsection{Model-based Policy Optimization}\label{sec:exp_model based}

Due to limitations in computational and memory resources, we consider a simple model-based scenario in this part. The MMS consists of three microgrids, each equipped with storage units but no demand load unit. Only Microgrid 1 has a renewable generator unit and can abandon power.

\begin{figure}[H]
\centering
\caption{The convergence procedure of Algorithm 1 under different values of $\beta$.}
\includegraphics[width=14cm]{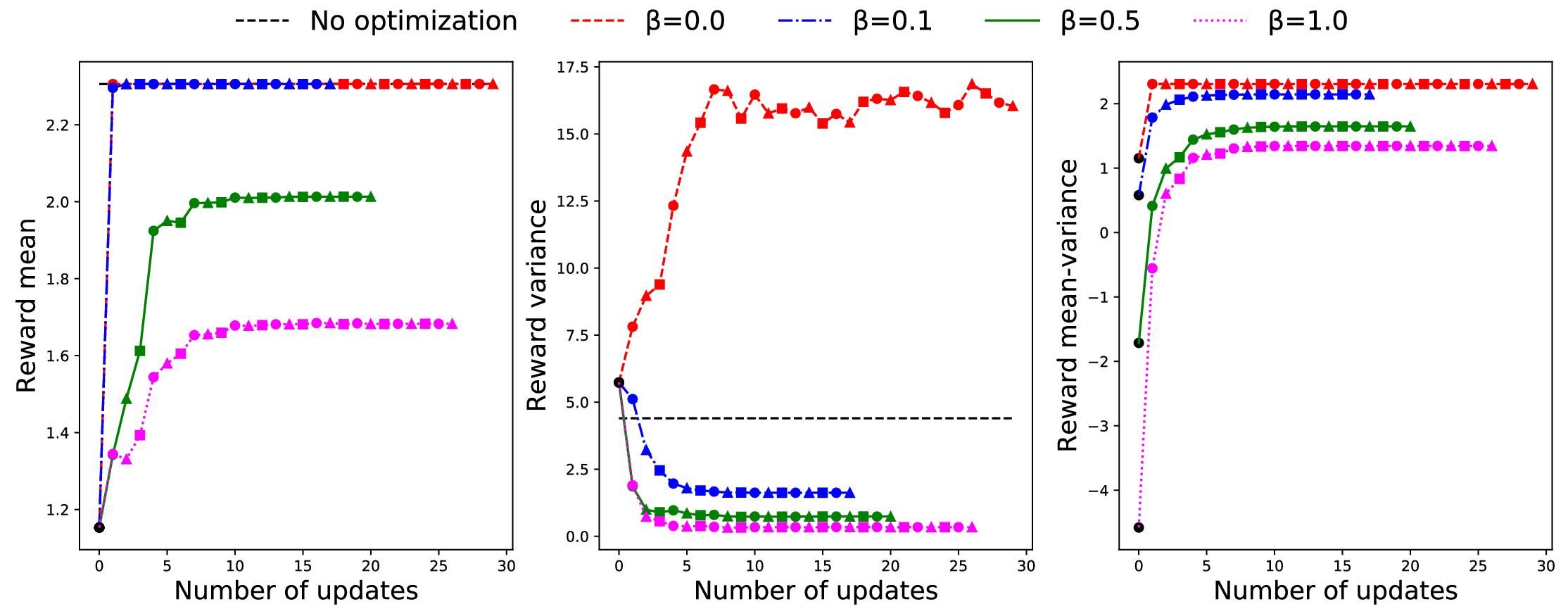}
\label{fig:converge_betas_scenario1}
\end{figure}

We evaluate Algorithm~\ref{alg:MVMAPI} with different coefficients $\beta$ and illustrate in Figure~\ref{fig:converge_betas_scenario1} the convergence of the reward mean, variance, and mean-variance. The initial policies and the update order remain fixed in all experiments. Three different point shapes on the curves represent the policy updates of three different microgrids. The black dashed lines in the first two sub-figures correspond to the results when no energy management is applied. The varying lengths of the curves indicate that the algorithm converges after different numbers of updates for different coefficients $\beta$.

When $\beta=0$, the algorithm optimizes only the mean value of the exchanged power, which aligns with the black dashed line after Microgrid 1 updates its policy. This alignment occurs because the long-term average exchanged power is not affected by the behavior of storage units and is solely determined by the power curtailment of Microgrid 1.

\huEdit{For $\beta=0.1$, the mean curve still converges to the maximum, while the variance decreases. These results suggest that when $\beta$ is small, microgrids can reduce power fluctuations by only controlling the storage units without energy abandonment. As $\beta$ increases, the variance of exchanged power is given more consideration and power curtailments are adopted by Microgrid 1, resulting in a reduction in the converged mean value. The third sub-figure demonstrates the monotonicity of Algorithm~\ref{alg:MVMAPI} when optimizing the mean-variance performance. }

\begin{figure}[!htbp]
    \small
    \centering
    \subfigure[Different initial policies]{
    \label{fig:converge_policy1}\includegraphics[width=0.4\columnwidth]{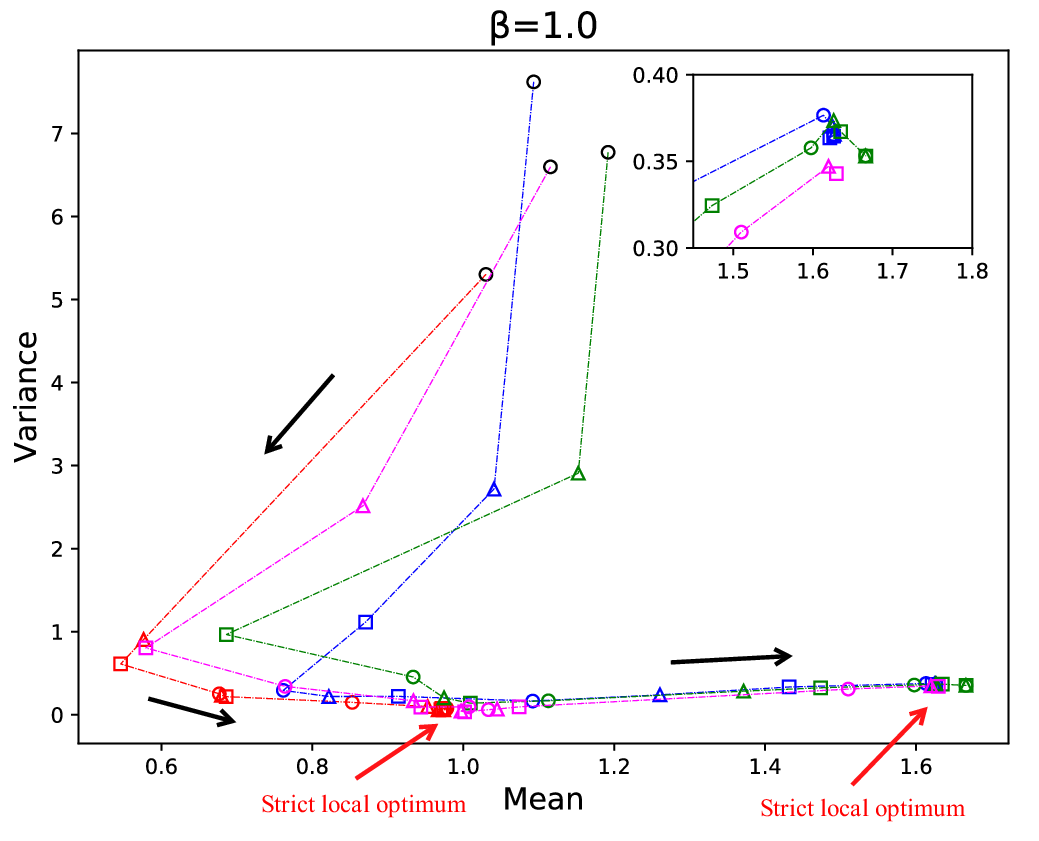} 
    }
    \subfigure[Different update orders]{
    \label{fig:converge_seq1}\includegraphics[width=0.4\columnwidth]{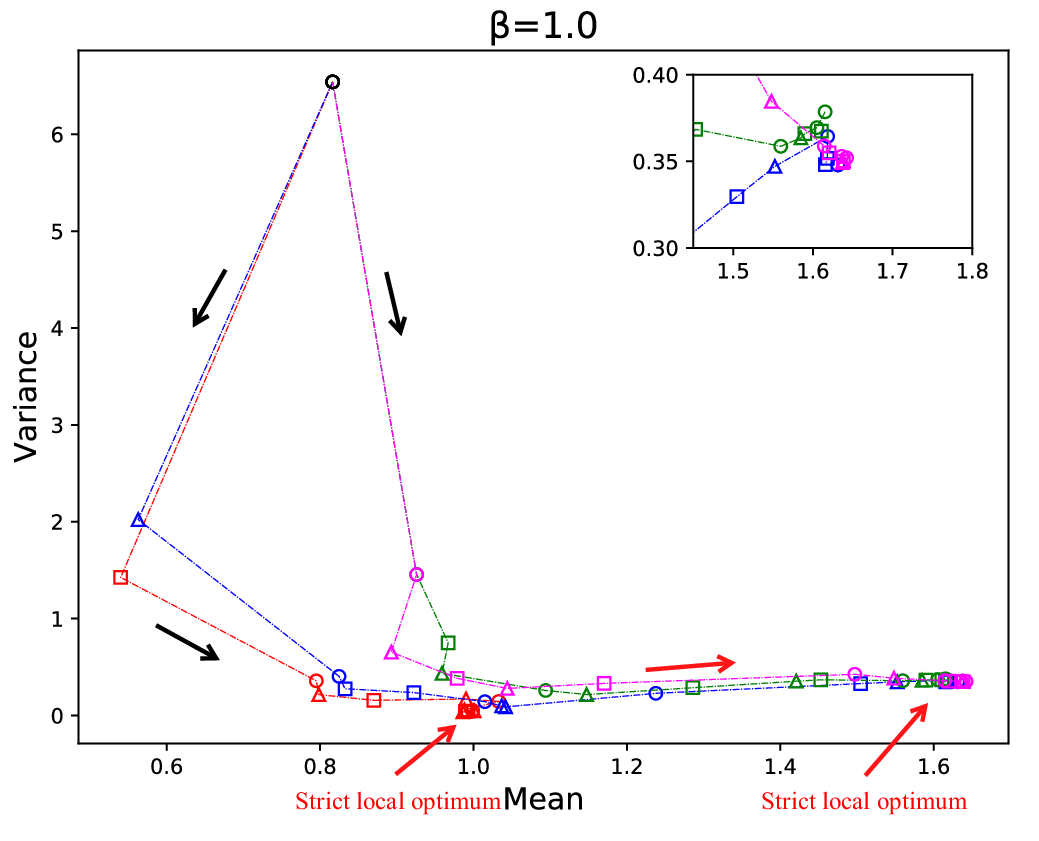} 
    }
    \caption{The convergence results under different initial policies or update orders when $\beta=1.0$.}
    \label{fig:converge_policyseq1}
\end{figure}

\huEdit{Next, we investigate how initial policies and update orders affect the converged value of Algorithm~\ref{alg:MVMAPI}. Figure~\ref{fig:converge_policyseq1} presents results for $\beta=1$. The first subfigure shows convergence from four different initial joint policies under a fixed update order, while the second illustrates results with a fixed initial policy but randomly generated update orders. }

As shown in Figure~\ref{fig:converge_policyseq1}, Algorithm~\ref{alg:MVMAPI} converges to different stationary points, which are verified to be strict local optima, due to variations in the initial joint policies and update orders. Since the optimal joint policy is first-order stationary, we can repeatedly apply MV-MAPI with various initial joint policies and update orders. The joint policy with the best performance is more likely to be the optimal solution.

\subsection{Multi-Agent Reinforcement Learning}\label{sec:exp_MARL}

\huEdit{We first evaluate MV-MATRPO in the MMS scenario of Section~\ref{sec:exp_model based} to compare its performance with MV-MAPI and verify its validity. We then test MV-MATRPO in a larger-scale MMS comprising five microgrids, each with a renewable generator, a demand load unit, and a storage unit. The parameters of Algorithm~\ref{alg:MVTRPO} are presented in Appendix~\ref{sec:appd_hyperparameter}.}

\begin{figure}[H]
    \centering
    \includegraphics[width=16cm]{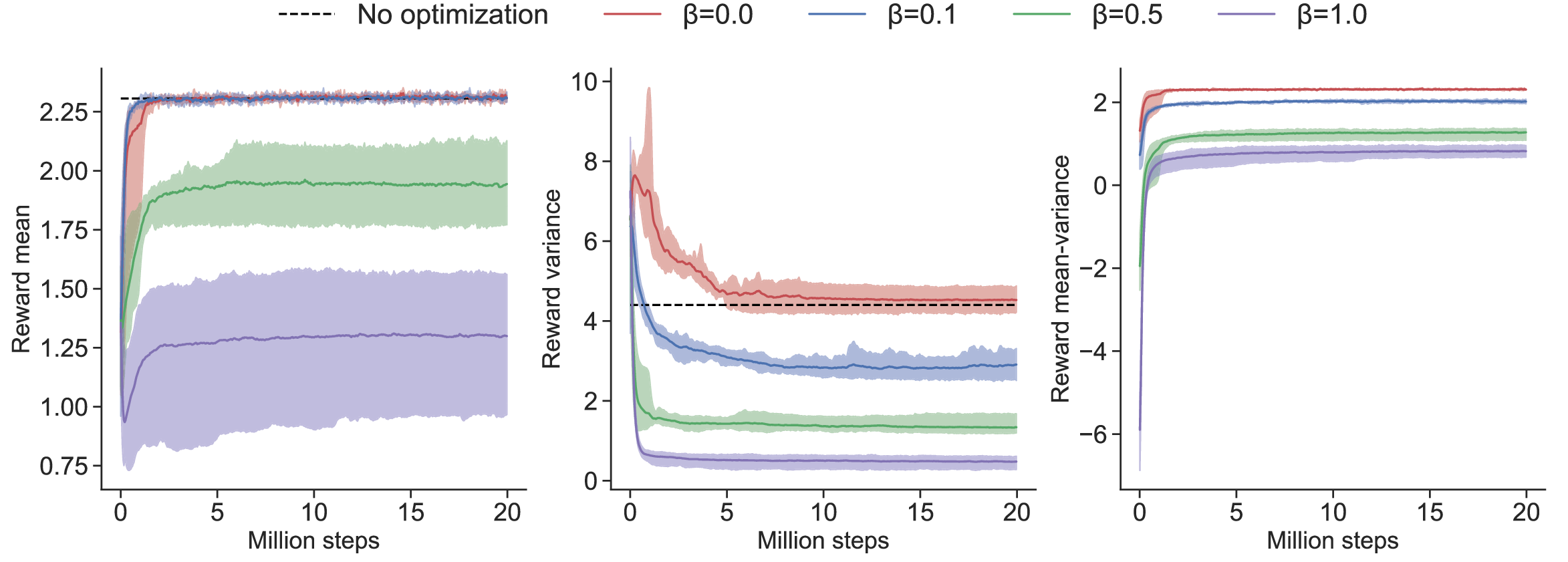}
    \caption{Scenario 1: the training curves of Algorithm 2 under different values of $\beta$.}
    \label{fig:tracurve-mg3}
    \end{figure}

\begin{figure}[H]
        \centering
        \includegraphics[width=16cm]{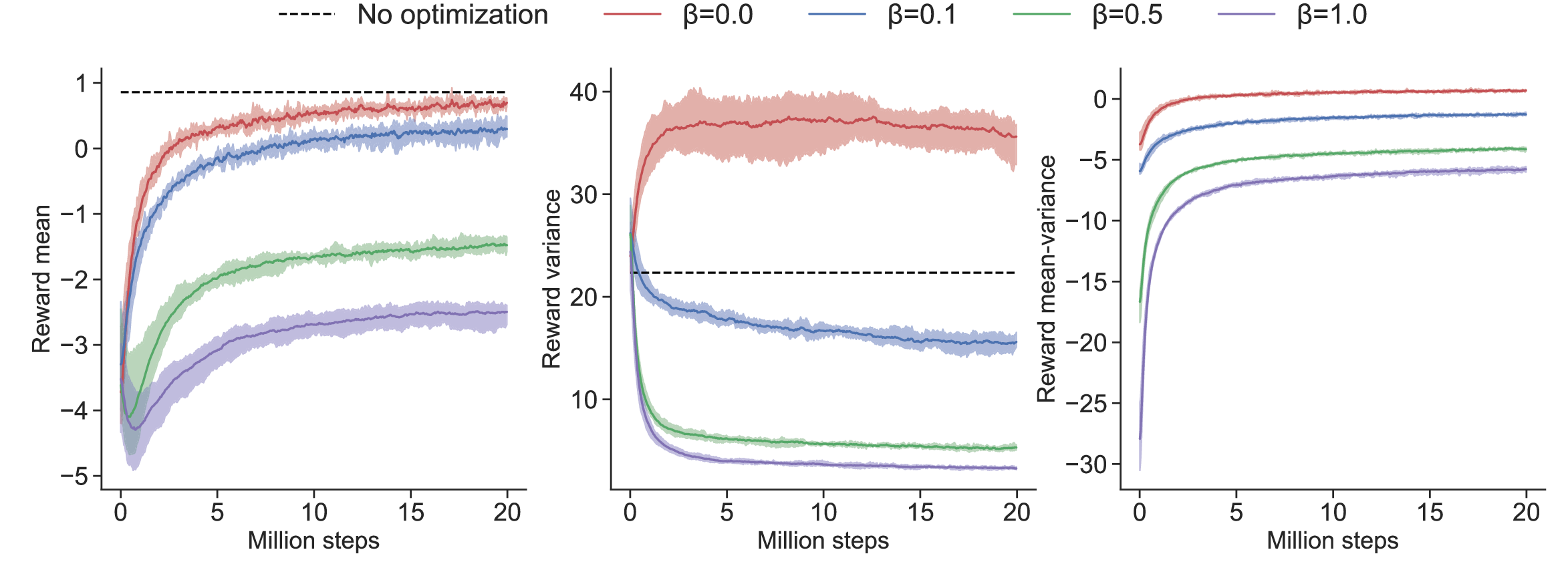}
        \caption{Scenario 2: the training curves of Algorithm 2  under different values of $\beta$.}
        \label{fig:tracurve-mg5}
\end{figure}

\huEdit{Experimental results are presented in Figures~\ref{fig:tracurve-mg3} and \ref{fig:tracurve-mg5}, where each training curve is averaged over six random seeds and shaded by the standard deviation. Initial policies and update orders are randomly generated. }


As Figure~\ref{fig:tracurve-mg3} illustrates, the training curves of the mean reward converge to the maximum when $\beta=0$ and $\beta=0.1$. And as $\beta$ increases, the variances are further reduced, accompanied by a decrease in the mean. The last sub-figure illustrates the monotonicity property of MV-MATRPO. These results are consistent with the observations in Figure~\ref{fig:converge_betas_scenario1}. 

Notably, in Figure~\ref{fig:tracurve-mg3}, when $\beta$ equals 0.5 and 1.0, the shaded regions of the converged curves of the reward mean are stable and wide. These results may be attributed to the fact that MV-MATRPO converges to different stationary points with different random seeds. For example, when $\beta=1.0$, the minimum and maximum of the converged reward mean are 0.97 and 1.56, respectively, which are very close to the results shown in Figure~\ref{fig:converge_policyseq1}.

Similar results are illustrated in Figure~\ref{fig:tracurve-mg5}, which demonstrates that even as the scenario becomes more complex, the algorithm can achieve a trade-off between the mean and variance with different coefficients $\beta$.

\section{Conclusion}\label{sec:conclusion}

In this paper, we study the long-run MV-TSG problem. The non-additive and non-Markovian nature of variance and the environmental non-stationarity in multi-agent settings render dynamic programming inapplicable. We address this problem through sensitivity-based optimization theory, deriving performance difference and performance derivative formulas in MV-TSGs. Subsequently, we develop the MV-MAPI algorithm by introducing a sequential update scheme. We prove that MV-MAPI converges monotonically to a first-order stationary point. Furthermore, we characterize the local geometry of these stationary points and provide verifiable conditions under which such points are (local) Nash equilibria or strict local optima in MV-TSGs. To address large-scale MV-TSGs in cases where environmental models are unknown, we propose the MV-MATRPO MARL algorithm. The proposed algorithms are evaluated on an MMS energy management problem, and experimental results validate both our main findings and the algorithms’ effectiveness.

One natural extension is to incorporate recurrent neural networks into MV-MATRPO
to approximately address TSGs with partial observability. Additionally, investigating other
methods to tackle the non-stationarity problem in stochastic games and developing algorithms
ensuring convergence to the global optimum are meaningful but challenging tasks.

\begin{spacing}{1.3}
\bibliographystyle{apalike} 
\bibliography{ref}
\end{spacing}
\vspace{2cm}
\newpage

\section*{Appendix}
\appendix
\section{Proofs}

\subsection{Proof of Lemma~\ref{lemma:J-difference}}
\label{sec:appd_J-difference}
\proof{}
    We introduce a pseudo mean $\omega$ to decompose the policy performance with policy-dependent reward. With $\omega$,   the original MV-TSG is transformed into a standard TSG with a reward function
    \begin{equation*}
        f_{\omega}(s,\bm{a}) = r(s,\bm{a})- \beta(r(s,\bm{a})-\omega)^2, \quad \forall s \in \mathcal{S}, \bm{a} \in \mathcal{A}.
    \end{equation*}
So the optimizing objective of the pseudo-reward function is given as 
    \begin{equation*}
        J_{\omega}^{\bm\mu}=\mathbb{E}_{s\sim \pi^{\bm\mu}, \bm{a} \sim \bm{\mu}}[f_\omega(s,\bm{a})].
    \end{equation*}
Since the pseudo reward is independent of the joint policy, the corresponding performance difference formula can be obtained directly according to \citet[Chapter 2]{cao2007stochastic}:
    \begin{equation}\label{eq:diff_pseudo}
        J_{\omega}^{\bm\mu'} - J_{\omega}^{\bm\mu}=\mathbb{E}_{s\sim \pi^{\bm\mu'}, \bm{a} \sim \bm\mu'}[A_{f_\omega}^{{\bm\mu}}(s,\bm{a})],
    \end{equation}
    where $A_{f_\omega}^{\bm\mu}(s,\bm{a})$ is the pseudo advantage with $f_\omega$ as the reward function. Subsequently, the performance difference formula of $J$ is derived as follows: 
    \begin{align}
        J(\bm\mu')-J(\bm\mu) &= (J_{\omega}^{\bm\mu'}-J_{\omega}^{\bm\mu}) + (J^{\bm\mu'}-J_{\omega}^{\bm\mu'}) + (J_{\omega}^{\bm\mu}-J^{\bm\mu}) \notag\\
            & = \mathbb{E}_{s\sim \pi^{\bm\mu'}, \bm{a} \sim \bm\mu'}[A_{f_\omega}^{\bm\mu}(s,\bm{a})] - \beta \mathbb{E}_{s \sim \pi^{\bm\mu'}, \bm{a} \sim \bm\mu'} [(r(s,\bm{a})-\eta^{\bm\mu'})^2 - (r(s,\bm{a})-\omega)^2] \notag\\
            & \quad -\beta \mathbb{E}_{s \sim \pi^{\bm\mu}, \bm{a} \sim \bm\mu} [(r(s,\bm{a})-\omega)^2 - (r(s,\bm{a})-\eta^{\bm\mu})^2]. \notag
    \end{align}
    Finally, by letting $\omega=\eta^{\bm\mu}$, we arrive at 
    \begin{equation}
        J(\bm\mu') - J(\bm\mu)=\mathbb{E}_{s \sim \pi^{\bm\mu'}, \bm{a} \sim \bm\mu'}[A_{f}^{\bm\mu}(s,\bm{a})] - \beta \mathbb{E}_{s \sim \pi^{\bm\mu'}, \bm{a} \sim \bm\mu'} [(r(s,\bm{a})-\eta^{\bm\mu'})^2 - (r(s,\bm{a})-\eta^{\bm\mu})^2].
        \label{eq:prf_MVPDF-1}
    \end{equation}

    Equation (\ref{eq:prf_MVPDF-1}) consists of two terms. The first term is associated with a standard TSG with $f$ as the reward function. The second term is caused by the perturbation of the mean and it can be further derived as
    \begin{align}
        &\beta \mathbb{E}_{s \sim \pi^{\bm\mu'}, \bm{a} \sim \bm\mu'} [(r(s,\bm{a})-\eta^{\bm\mu'})^2 - (r(s,\bm{a})-\eta^{\bm\mu})^2] \notag\\
        &= \beta\sum\limits_s \pi^{\bm\mu'}(s) \sum\limits_{\bm{a}} \bm\mu'(\bm{a}|s) [(\eta^{\bm\mu'})^2-2r(s,\bm{a}) \eta^{\bm\mu'} +2r(s,\bm{a})\eta^{\bm\mu} - (\eta^{\bm\mu})^2] \notag\\
        &= \beta((\eta^{\bm\mu'})^2 - 2(\eta^{\bm\mu'})^2 + 2\eta^{\bm\mu} \eta^{\bm\mu'} - (\eta^{\bm\mu})^2) \notag\\
        &= - \beta(\eta^{\bm\mu'}-\eta^{\bm\mu})^2, \label{eq:prf_MVPDF-2}
    \end{align}
    where the second derivation uses the result $\sum\limits_s \pi^{\bm\mu'}(s) \sum\limits_{\bm{a}} \bm\mu'(\bm{a}|s) r(s,\bm{a}) =\eta^{\bm\mu'}$. Substituting (\ref{eq:prf_MVPDF-2}) into  (\ref{eq:prf_MVPDF-1}), we obtain (\ref{eq:MVPDF-2}).
\endproof

\subsection{Proof of Lemma~\ref{lemma:J-derivative}}
\label{sec:appd_J-derivative}
\proof{}
    By Lemma~\ref{lemma:J-difference}, the difference between $\bm\mu$ and $\delta_{\bm\mu}^{\bm\mu'}$ is
    \begin{equation*}
        J(\delta_{\bm\mu}^{\bm\mu'}) - J(\bm\mu) =\mathbb{E}_{s \sim \pi^{\delta_{\bm\mu}^{\bm\mu'}}, \bm{a} \sim \delta_{\bm\mu}^{\bm\mu'}}[A_{f}^{\bm\mu} (s,\bm{a})] - \beta \mathbb{E}_{s \sim \pi^{\delta_{\bm\mu}^{\bm\mu'}}, \bm{a} \sim \delta_{\bm\mu}^{\bm\mu'}} [(r(s,\bm{a})-\eta^{\delta_{\bm\mu}^{\bm\mu'}})^2 - (r(s,\bm{a})-\eta^{\bm\mu})^2].
    \end{equation*}
The performance derivative formula can be obtained by taking the derivative w.r.t. $\delta$ and letting $\delta \rightarrow 0$.  Denote 
\begin{align*}
    & l_1(\delta) = \mathbb{E}_{s \sim \pi^{\delta_{\bm\mu}^{\bm\mu'}}, \bm{a} \sim \delta_{\bm\mu}^{\bm\mu'}}[A_{f}^{\bm\mu} (s,\bm{a})], \\
    & l_2(\delta) = \mathbb{E}_{s \sim \pi^{\delta_{\bm\mu}^{\bm\mu'}}, \bm{a} \sim \delta_{\bm\mu}^{\bm\mu'}} [(r(s,\bm{a})-\eta^{\delta_{\bm\mu}^{\bm\mu'}})^2 - (r(s,\bm{a})-\eta^{\bm\mu})^2].
\end{align*}
Then $J(\delta_{\bm\mu}^{\bm\mu'}) - J(\bm\mu) = l_1(\delta)-\beta l_2(\delta)$. For $l_1(\delta)$,
    \begin{align*}
        l_1(\delta) &= \mathbb{E}_{s \sim \pi^{\delta_{\bm\mu}^{\bm\mu'}}} \Big\{(1-\delta) \mathbb{E}_{\bm{a} \sim \bm\mu}[A_{f}^{\bm\mu}(s,\bm{a})] + \delta \mathbb{E}_{\bm{a} \sim \bm\mu'}[A_{f}^{\bm\mu}(s,\bm{a})] \Big\} \\
                    &= \delta \mathbb{E}_{s \sim \pi^{\delta_{\bm\mu}^{\bm\mu'}}, \bm{a} \sim \bm\mu'}[A_{f}^{\bm\mu}(s,\bm{a})] ,
    \end{align*}
    where the last equality follows that $\mathbb{E}_{\bm{a} \sim \bm\mu}[A_{f}^{\bm\mu}(s,\bm{a})]=0$. Since $\lim_{\delta \rightarrow 0} \delta_{\bm\mu}^{\bm\mu'} =\bm\mu$, it follows that
    \begin{equation*}
        \frac{\text{d}l_1(\delta)}{\text{d}\delta} \Big|_{\delta=0}=\mathbb{E}_{s \sim \pi^{\bm\mu}, \bm{a} \sim \bm\mu'}[A_{f}^{\bm\mu}(s,\bm{a})].  
    \end{equation*}

    For $l_2(\delta)$, 
    \begin{align*}
        \frac{\text{d}l_2(\delta)}{\text{d}\delta}\Big|_{\delta=0} 
        &= \lim\limits_{\delta \rightarrow 0} \frac{l_2(\delta)-l_2(0)}{\delta} \\
        &=\lim\limits_{\delta \rightarrow 0} \frac{1}{\delta} \sum\limits_s \pi^{\delta_{\bm\mu}^{\bm\mu'}}(s) \sum\limits_{\bm{a}} \delta_{\bm\mu}^{\bm\mu'}(\bm{a}|s)[(r(s,\bm{a})-\eta^{\delta_{\bm\mu}^{\bm\mu'}})^2 - (r(s,\bm{a}) - \eta^{\bm\mu})^2] \\
        &= \sum\limits_s \pi^{\bm\mu}(s) \sum\limits_{\bm{a}} \bm\mu(\bm{a}|s) \frac{\text{d}(r(s,\bm{a})-\eta^{\delta_{\bm\mu}^{\bm\mu'}})^2}{\text{d}\delta} \Big|_{\delta=0} \\
        &= \sum\limits_s \pi^{\bm\mu}(s) \sum\limits_{\bm{a}} \bm\mu(\bm{a}|s) \Big[-2(r(s,\bm{a})-\eta^{\bm\mu})\frac{\text{d}\eta^{\delta_{\bm\mu}^{\bm\mu'}}}{\text{d}\delta}\Big|_{\delta=0} \Big] \\
        &=0,
    \end{align*}
    where the last equality follows that $\sum\limits_s \pi^{\bm\mu}(s) \sum\limits_{\bm{a}}  \bm\mu(\bm{a}|s)r(s,\bm{a})=\eta^{\bm\mu}$, $\sum\limits_s \pi^{\bm\mu}(s)=1$ and $\sum\limits_{\bm{a}} \bm\mu(\bm{a}|s) =1$. Therefore,
    \begin{align*}
        \frac{\text{d}J(\delta_{\bm\mu}^{\bm\mu'})}{\text{d}\delta} 
        &= \frac{\text{d}l_1(\delta)}{\text{d}\delta} - \beta \frac{\text{d}l_2(\delta)}{\text{d}\delta} \\
        &= \mathbb{E}_{s\sim \pi^{\bm\mu}, \bm{a} \sim \bm\mu'}[A_{f}^{\bm\mu}(s,\bm{a})].
    \end{align*}
    
    The above equality indicates that the performance derivative is related to the surrogate reward function $f^{\bm\mu}(s,\bm{a})=r(s,\bm{a}) -\beta (r(s,\bm{a})-\eta^{\bm\mu})^2$.
\endproof

\subsection{Proof of Theorem~\ref{threm:deterministic NE}}
\label{sec:appd_threm1}
\proof{}
    According to the definition, it is evident that the global optimal joint policy of the MV-TSG is an NE. We now show that there exists a global optimal joint policy that is deterministic. \cite{xia2020risk} has proven that a deterministic policy can achieve the optimum in mean-variance MDPs.
    Accordingly, there exists a deterministic centralized joint policy
    \begin{equation*}
        \bm{\mu}^*(\bm{a}|s) = \bm{1} \{\bm{a}=(a^*_1(s),\ldots,a^*_N(s))  \}, \quad s\in \mathcal{S}    \end{equation*}
    that satisfies
    \begin{equation*}
        \bm{\mu}^*=\mathop{\arg\max}\limits_{\bm{\mu}} J^{\bm{\mu}}.
    \end{equation*}
    Let $\mu^*_i(a_i|s)=\bm{1} \{a_i=a^*_i(s) \},  s\in \mathcal{S}$ , we have
    \begin{equation*}
        \bm{\mu}^*(\bm{a}|s)=\prod_{i=1}^N \mu^*_i (a_i|s), \quad  s\in \mathcal{S}.
    \end{equation*}
    Since $\bm{\mu}^*$ globally maximizes the mean-variance performance function $J^{\bm\mu}$, the distributed deterministic policies $(\mu^*_1,\ldots,\mu^*_N)$ also maximize $J^{\bm\mu}$ globally, which completes the proof.
\endproof

\subsection{Proof of Theorem~\ref{theorem:stationary policy}}
\label{sec:appd_stationary policy}
\proof{}

At the stationary point $\tilde{\bm\mu}$, if for some agent $i$, 
$\frac{\text{d}J(\delta_{\tilde{\mu}_i}^{\mu_i'},\tilde{\bm\mu}_{-i})}{\text{d}\delta}\Big|_{\delta=0} < 0$ holds along the direction of any other policy $\mu_i'$, then there exists $ \bar\delta_i \in (0,1]$ such that for all $ \delta \in (0,\bar\delta_i]$, $J(\delta_{\tilde{\mu}_i}^{\mu_i'},\tilde{\mu}_{-i}) < J(\tilde{\mu}_i,\tilde{\mu}_{-i})$. If  Inequality~(\ref{eq:stationay_neq}) holds strictly for every agent, a common constant $\bar\delta=\min(\bar\delta_1,\ldots,\bar\delta_N)$ can be chosen. According to  Definition~\ref{def:local_NE}, the first-order stationary point is therefore a strict local NE, which completes the proof of the first statement of Theorem~\ref{theorem:stationary policy}.

However, if the first-order stationary point $\tilde{\bm\mu}$ satisfies the condition that there exists some agent $i$ and policy $\mu_i'$ such that  $\frac{\text{d}J(\delta_{\tilde{\mu}_i}^{\mu_i'},\tilde{\bm\mu}_{-i})}{\text{d}\delta}\Big|_{\delta=0} = 0$, the analysis of such cases becomes more intricate. 
When $\frac{\text{d}J(\delta_{\tilde{\mu}_i}^{\mu_i'},\tilde{\bm\mu}_{-i})}{\text{d}\delta}\Big|_{\delta=0} = 0$, according to Lemma~\ref{lemma:J-derivative}, it follows that
\begin{equation*}
    \frac{\text{d}J(\delta_{\tilde{\mu}_i}^{\mu_i'},\tilde{\bm\mu}_{-i})}{\text{d}\delta}\Big|_{\delta=0} = \mathbb{E}_{s\sim \pi^{\tilde{\bm\mu}}, {a_i\sim \mu_i', \bm{a}_{-i}\sim \tilde{\bm\mu}_{-i}}}[A_{f}^{\tilde{\bm\mu}}(s,a_i, \bm{a}_{-i})]=0.
\end{equation*}
Because the elements of steady-state distribution $\pi^{\tilde{\bm\mu}}$ are positive, 
and $\frac{\text{d}J(\delta_{\tilde{\mu}_i}^{\mu_i},\tilde{\bm\mu}_{-i})}{\text{d}\delta}\Big|_{\delta=0} \le 0$ holds for any agent $i$ and policy direction $\mu_i \in \mathcal{U}_i$, 
we claim that
\begin{equation}
    \mathbb{E}_{a_i\sim \mu_i', \bm{a}_{-i}\sim \tilde{\bm\mu}_{-i}} [A_{f}^{\tilde{\mu}_i,\tilde{\bm\mu}_{-i}}(s,a_i,\bm{a}_{-i})] = 0, \forall s \in \mathcal{S}.
    \label{eq:key_eq}
\end{equation}

This claim is proved by contradiction. If for some state $s$, $\mathbb{E}_{a_i\sim \mu_i', \bm{a}_{-i}\sim \tilde{\bm\mu}_{-i}} [A_{f}^{\tilde{\mu}_i,\tilde{\bm\mu}_{-i}}(s,a_i,\bm{a}_{-i})] > 0$, a new policy $\mu_i''$ can be constructed such that $\mu_i''(a_i|s)=\mu'_i(a_i|s),\forall a_i \in \mathcal{A}_i$, and $\mu_i''(\cdot|s)=\tilde{\mu}_i(\cdot|s)$ in all other states. Let $\delta_{\tilde{\mu}_i}^{\mu_i''} = (1-\delta)\tilde{\mu}_i + \delta \mu_i''$, then $\frac{\text{d}J(\delta_{\tilde{\mu}_i}^{\mu''_i},\tilde{\bm\mu}_{-i})}{\text{d}\delta}\Big|_{\delta=0} >0$, which contradicts with the fact that $\tilde{\bm\mu}$ is first-order stationary.

According to the performance difference of Lemma~\ref{lemma:J-difference}, it follows that 
\begin{align}
    J(\delta_{\tilde{\mu}_i}^{\mu'_i},\tilde{\bm\mu}_{-i}) - J(\tilde{\mu}_i,\tilde{\bm\mu}_{-i}) 
    =& \mathbb{E}_{s\sim \pi^{\delta_{\tilde{\mu}_i}^{\mu_i'},\tilde{\bm\mu}_{-i}}, \bm{a}\sim (\delta_{\tilde{\mu}_i}^{\mu'_i},\tilde{\bm\mu}_{-i})} [A_{f}^{\tilde{\mu}_i,\tilde{\bm\mu}_{-i}}(s,\bm{a})] + \beta(\eta^{\delta_{\tilde{\mu}_i}^{\mu_i'},\tilde{\bm\mu}_{-i}} - \eta^{\tilde{\mu}_i,\tilde{\bm\mu}_{-i}})^2 \notag\\
    =& (1-\delta)\mathbb{E}_{s\sim \pi^{\delta_{\tilde{\mu}_i}^{\mu'_i},\tilde{\bm\mu}_{-i}}, \bm{a}\sim (\tilde{\mu}_i,\tilde{\bm\mu}_{-i})} [A_{f}^{\tilde{\mu}_i,\tilde{\bm\mu}_{-i}}(s,\bm{a})] \notag\\
    & + \delta \mathbb{E}_{s\sim \pi^{\delta_{\tilde{\mu}_i}^{\mu'_i},\tilde{\bm\mu}_{-i}}, \bm{a}\sim (\mu'_i,\tilde{\bm\mu}_{-i})} [A_{f}^{\tilde{\mu}_i,\tilde{\bm\mu}_{-i}}(s,\bm{a})] \notag\\
    & + \beta(\eta^{\delta_{\tilde{\mu}_i}^{\mu'_i},\tilde{\bm\mu}_{-i}} - \eta^{\tilde{\mu}_i,\tilde{\bm\mu}_{-i}})^2  \notag\\
     =& \beta(\eta^{\delta_{\tilde{\mu}_i}^{\mu'_i},\tilde{\bm\mu}_{-i}} - \eta^{\tilde{\mu}_i,\tilde{\bm\mu}_{-i}})^2. \label{eq:perf_diff_stationary}
\end{align}
The second equality holds because  $\delta_{\tilde{\mu}_i}^{\mu'_i}(s,a_i)=(1-\delta)\tilde{\mu}(s,a_i) + \delta \mu'(s,a_i)$. The third equality is due to $\mathbb{E}_{\bm{a}\sim \tilde{\bm\mu}} [A_{f}^{\tilde{\bm\mu}}(s,\bm{a})]=0$ and $\mathbb{E}_{a_i \sim \mu'_i, \bm{a}_{-i}\sim \tilde{\mu}_{-i}} [A_{f}^{\tilde{\mu}_i,\tilde{\bm\mu}_{-i}}(s,a_i,\bm{a}_{-i})]=0$ (Equation~(\ref{eq:key_eq})). Equation (\ref{eq:perf_diff_stationary}) indicates that for arbitrarily small $\delta$, if $\eta^{\delta_{\tilde{\mu}_i}^{\mu'_i},\tilde{\bm\mu}_{-i}} \neq \eta^{\tilde{\mu}_i,\tilde{\bm\mu}_{-i}}$, then  $J(\delta_{\tilde{\mu}_i}^{\mu'_i},\tilde{\bm\mu}_{-i}) > J(\tilde{\mu}_i,\tilde{\bm\mu}_{-i})$. Therefore, when $\frac{\text{d}J(\delta_{\tilde{\mu}_i}^{\mu'_i},\tilde{\bm\mu}_{-i})}{\text{d}\delta}\Big|_{\delta=0} = 0$ along the direction of some policy $\mu'_i$, the necessary and sufficient  condition for $\tilde{\mu}_i$ to be local optimal in the direction of $\mu'_i$ is that $\exists \bar\delta_i \in (0,1], \forall \delta \in (0,\bar\delta_i]$, we have $\eta^{\delta_{\tilde{\mu}_i}^{\mu'_i},\tilde{\bm\mu}_{-i}} = \eta^{\tilde{\mu}_i,\tilde{\bm\mu}_{-i}}$. If the necessary and sufficient  condition holds for any agent $i$ and any policy $\mu'_i$ satisfying $\frac{\text{d}J(\delta_{\tilde{\mu}_i}^{\mu'_i},\tilde{\bm\mu}_{-i})}{\text{d}\delta}\Big|_{\delta=0} = 0$, then the first-order stationary point is a local NE. Furthermore, applying (\ref{eq:perf_diff_stationary}) with $\delta=1$ directly leads to Corollary~\ref{coro:escape_saddle}.
\endproof

\subsection{Proof of Theorem~\ref{theorem:strict local optima}}
\label{sec:app_ne_localoptimal}
\proof{}
We assume that a strict local Nash joint policy is $\bm\mu^*=(\mu_1^*,\ldots,\mu_N^*)$ and any other joint policy is $\bm\mu'=(\mu'_1,\ldots,\mu'_N)$. Let $\delta_{\bm\mu^*}^{\bm\mu'}=(1-\delta)\bm\mu^* + \delta \bm\mu'$ and $\Delta_i = \mu'_i - \mu_i^*$, then
\begin{align*}
    \frac{\text{d}J(\delta_{\bm\mu^*}^{\bm\mu'})}{\text{d}\delta}\Big|_{\delta=0} 
    &= \lim\limits_{\delta \to 0} \frac{J(\mu_1^*+\delta \Delta_1, \ldots, \mu_N^*+\delta \Delta_N) - J(\mu_1^*,\ldots,\mu_N^*)}{\delta} \\
    &= \sum\limits_{i=1}^N \frac{\partial J(\mu_i,\mu_{-i}^*)}{\partial \mu_i}\Big|_{\mu_i=\mu_i^*} \frac{\Delta_i}{|\Delta \bm\mu|} \\
    &< 0.
\end{align*}
The second equality holds for the calculation of directional derivative, and $|\Delta \bm\mu|$ indicates the magnitude of $\Delta \bm\mu$ (across all state-action pairs). The inequality holds because strict local Nash joint policies satisfy $\frac{\partial J(\mu_i,\mu_{-i}^*)}{\partial \mu_i}\Big|_{\mu_i=\mu_i^*} \frac{\Delta_i}{|\Delta_i|} < 0$ ( $|\Delta_i| \neq 0$) for all agents $i$. Thus, strict local NEs are strict local optima in MV-TSGs. 

Next, we demonstrate that the converse also holds. For a strict local optimal joint policy $\bm\mu^*$ and any mixed joint policy $\delta_{\bm\mu^*}^{\bm\mu}=(1-\delta)\bm\mu^* + \delta \bm\mu, \bm\mu\in \mathcal{U}$, we have $\frac{\text{d}J(\delta_{\bm\mu^*}^{\bm\mu})}{\text{d}\delta}\Big|_{\delta=0} < 0$. 
Given a policy $\mu'_i$ and $\bm\mu' = (\mu'_i, \bm\mu_{-i}^*)$, a mixed joint policy is constructed by $\delta_{\bm\mu^*}^{\bm\mu'} = (\delta_{\mu_i^*}^{\mu'_i}, \bm\mu_{-i}^*) $. Due to $\frac{\text{d}J(\delta_{\bm\mu^*}^{\bm\mu'})}{\text{d}\delta}\Big|_{\delta=0}=\frac{\text{d}J(\delta_{\mu_i^*}^{\mu'_i}, \bm\mu_{-i}^*)}{\text{d}\delta}\Big|_{\delta=0}<0$ along the direction of any policy $\mu'_i \in \mathcal{U}_i$, strict local optimal joint policies are strict local NEs according to Definition~\ref{def:local_NE}. The proof is finished.
\endproof

\subsection{Proof of Proposition~\ref{prop:A_estimator}} \label{sec:apped_prop2}
\proof{}
\begin{align*}
    \mathbb{E}_{\bm{a}\sim \bm\mu}& \big[
        (\frac{\hat{\mu}_{i_h} (a_{i_h}|s)}{\mu_{i_h} (a_{i_h}|s )}-1)  \frac{\bm\mu'_{i_{1:h-1}}(\bm{a}_{i_{1:h-1}}|s)}{\boldsymbol{\mu}_{i_{1:h-1}}(\bm{a}_{i_{1:h-1}}|s)} A_{f}^{\bm\mu}(s,\bm{a})  \big] \\
        =& \mathbb{E}_{\bm{a}\sim \bm\mu} \big[
        (\frac{\hat{\mu}_{i_h} (a_{i_h}|s)}{\mu_{i_h} (a_{i_h}|s )})  \frac{\bm\mu'_{i_{1:h-1}}(\bm{a}_{i_{1:h-1}}|s)}{\boldsymbol{\mu}_{i_{1:h-1}}(\bm{a}_{i_{1:h-1}}|s)} A_{f}^{\bm\mu}(s,\bm{a})) -\frac{\bm\mu'_{i_{1:h-1}}(\bm{a}_{i_{1:h-1}}|s)}{\boldsymbol{\mu}_{i_{1:h-1}}(\bm{a}_{i_{1:h-1}}|s)} A_{f}^{\bm\mu}(s,\bm{a}) \big] \\
        =& \mathbb{E}_{\bm{a}_{i_{1:h}}\sim \bm\mu_{i_{1:h}}, \bm{a}_{-i_{1:h}}\sim \bm\mu_{-i_{1:h}}} \big[
        \big(\frac{\hat{\mu}_{i_h} (a_{i_h}|s)}{\mu_{i_h} (a_{i_h}|s )})  \frac{\bm\mu'_{i_{1:h-1}}(\bm{a}_{i_{1:h-1}}|s)}{\boldsymbol{\mu}_{i_{1:h-1}}(\bm{a}_{i_{1:h-1}}|s)} A_{f}^{\bm\mu}(s,\bm{a}_{i_{1:h}}, \bm{a}_{-i_{1:h}})\big] \\
        & - \mathbb{E}_{\bm{a}_{i_{1:h-1}}\sim \bm\mu_{i_{1:h-1}}, \bm{a}_{-i_{1:h-1}}\sim \bm\mu_{-i_{1:h-1}}} \big[  \frac{\bm\mu'_{i_{1:h-1}}(\bm{a}_{i_{1:h-1}}|s)}{\boldsymbol{\mu}_{i_{1:h-1}}(\bm{a}_{i_{1:h-1}}|s)} A_{f}^{\bm\mu}(s,\bm{a}_{i_{1:h-1}}, \bm{a}_{-i_{1:h-1}})\big] \\
        =& \mathbb{E}_{\bm{a}_{i_{1:h-1}}\sim \bm\mu'_{i_{1:h-1}}, a_{i_h}\sim \hat{\mu}_i, \bm{a}_{-i_{1:h}}\sim \bm\mu_{-i_{1:h}}} \big[
         A_{f}^{\bm\mu}(s,\bm{a}_{i_{1:h}}, \bm{a}_{-i_{1:h}})\big] \\
         & -\mathbb{E}_{\bm{a}_{i_{1:h-1}}\sim \bm\mu'_{i_{1:h-1}}, \bm{a}_{-i_{1:h-1}}\sim \bm\mu_{-i_{1:h-1}}} \big[ A_{f}^{\bm\mu}(s,\bm{a}_{i_{1:h-1}}, \bm{a}_{-i_{1:h-1}})\big] \\
        =& \mathbb{E}_{\bm{a}_{i_{1:h-1}}\sim \bm\mu'_{i_{1:h-1}}, a_{i_h}\sim \hat{\mu}_{i_h}} \Big[ \mathbb{E}_{\bm{a}_{-i_{1:h}}\sim \bm\mu_{-i_{1:h}}} \big[
         A_{f}^{\bm\mu}(s,\bm{a}_{i_{1:h}}, \bm{a}_{-i_{1:h}})\big] \Big] \\
         &- \mathbb{E}_{\bm{a}_{i_{1:h-1}}\sim \bm\mu'_{i_{1:h-1}}} \Big[\mathbb{E}_{\bm{a}_{-i_{1:h-1}}\sim \bm\mu_{-i_{1:h-1}}} \big[ A_{f}^{\bm\mu}(s,\bm{a}_{i_{1:h-1}}, \bm{a}_{-i_{1:h-1}})\big] \Big]  \\
        =& \mathbb{E}_{\bm{a}_{i_{1:h-1}}\sim \bm\mu'_{i_{1:h-1}}, a_{i_h}\sim \hat{\mu}_i} \big[
         A_{f}^{\bm\mu}(s,\bm{a}_{i_{1:h}})\big] 
         - \mathbb{E}_{\bm{a}_{i_{1:h-1}}\sim \bm\mu'_{i_{1:h-1}}} \big[ A_{f}^{\bm\mu}(s,\bm{a}_{i_{1:h-1}})\big] \\
        =& \mathbb{E}_{\bm{a}_{i_{1:h-1}}\sim \bm\mu'_{i_{1:h-1}}, a_{i_h}\sim \hat{\mu}_i} \big[
         A_{f}^{\bm\mu}(s,\bm{a}_{i_{1:h-1}}, a_{i_h})\big].
\end{align*}
\endproof

\subsection{Proof of Theorem~\ref{threm:mv_bounds}}
\label{sec:app_joint_bound}
\proof{}
We start our analysis from the mean-variance performance difference formula 
\begin{equation*}
    J^{\bm{\mu}'} - J^{\bm{\mu}} 
        =\mathbb{E}_{s \sim \pi^{\bm{\mu}'}, \bm{a} \sim \bm{\mu}'}[A_{f}^{\bm{\mu}}(s,\bm{a})] + \beta (\eta^{\bm\mu'} -\eta^{\bm\mu})^2.
\end{equation*}
The result indicates that the performance difference can be decomposed into two terms. The first term, associated with the mean-variance reward function $f$, can be addressed using the standard average trust region method.  Lemma~\ref{lemma:trpo_joint} indicates that
\begin{equation}
\label{eq:mv_bounds_term1}
    \mathbb{E}_{s \sim \pi^{\bm\mu'}, \bm{a}\sim \bm\mu'}[A_{f}^{\bm\mu}(s,\bm{a})] - \mathcal{L}_{f}^{\bm\mu}(\bm\mu') 
    \ge - 2(\kappa^* - 1) \epsilon_f D_{\text{TV}}(\bm\mu',\bm\mu) .
\end{equation}
To bound the second term, we want to find a quantity $H\ge0$ such that 
\begin{equation*}
    (\eta^{\bm\mu'}-\eta^{\bm\mu})^2 \ge H^2. \notag
\end{equation*}
The square term can be lower bound by 0, or by the square of a lower bound of its argument if the latter is greater than 0. Due to its convexity, a square function attains a strictly positive minimum when either the upper bound of its argument is negative or the lower bound of its argument is positive.

\cite{zhang2021policy} demonstrate that the bounds of the average reward trust region method as
\begin{equation*}
    \mathcal{L}^{\bm\mu}(\bm\mu')-2(\kappa^*-1)\epsilon_{\eta}D_{\text{TV}}(\bm\mu',\bm\mu) \le \eta^{\bm\mu'}-\eta^{\bm\mu} \le \mathcal{L}^{\bm\mu}(\bm\mu')+ 2(\kappa^*-1)\epsilon_{\eta}D_{\text{TV}}(\bm\mu',\bm\mu),
\end{equation*}
where $\mathcal{L}^{\bm\mu}(\bm\mu')=\mathbb{E}_{s\sim \pi^{\bm{\mu}},\bm{a}\sim \bm\mu'}[A^{\bm\mu}(s,\bm{a})]$ and $\epsilon_{\eta}= \max\limits_s \mathbb{E}_{\bm{a}\sim \bm\mu'}[A^{\bm\mu}(s,\bm{a})]$ (see Lemma~\ref{lemma:trpo_joint}).

The best lower bound can be obtained by taking the maximum among the argument lower bound, the opposite of the argument upper bound and 0, i.e., $H=\max(0, \mathcal{L}^{\bm\mu}(\bm\mu')-2(\kappa^*-1)\epsilon_{\eta}D_{\text{TV}}(\bm\mu',\bm\mu), -\mathcal{L}^{\bm\mu}(\bm\mu')- 2(\kappa^*-1)\epsilon_{\eta}D_{\text{TV}}(\bm\mu',\bm\mu))$, finally taking the square of this quantity. Then, we arrive at
\begin{equation*}
    J^{\bm\mu'} - J^{\bm\mu} \ge \mathcal{L}_{f}^{\bm\mu}(\bm\mu') - 2(\kappa^*-1)\epsilon_f D_{\text{TV}}(\bm\mu',\bm\mu) + \beta H^2,
\end{equation*}
where $H =\max(0, \mathcal{L}^{\bm\mu}(\bm\mu')-2(\kappa^*-1)\epsilon_{\eta}D_{\text{TV}}(\bm\mu',\bm\mu), -\mathcal{L}^{\bm\mu}(\bm\mu')- 2(\kappa^*-1)\epsilon_{\eta}D_{\text{TV}}(\bm\mu',\bm\mu))$.
\endproof

\subsection{Proof of Theorem~\ref{threm:ma_bounds}}
\label{sec:app_ma_bound}
\proof{}
    We start this proof from Theorem~\ref{threm:mv_bounds}. Since the variable $H$ in Theorem~\ref{threm:mv_bounds} is difficult to consider and is always positive, we can bound the mean-variance performance by neglecting it. In this case, together with (\ref{eq:mv_bounds_term1}), it follows that
    \begin{equation}
        J^{\bm\mu'} - J^{\bm\mu} \ge \mathcal{L}_{f}^{\bm\mu}(\bm\mu') - 2(\kappa^*-1)\epsilon_f 
        \left[ \mathbb{E}_{s\sim \pi^{\bm\mu}} D_{\text{TV}}(\bm\mu'(\cdot|s) \| \bm\mu(\cdot|s))\right] .
        \label{eq:mv_bounds_bref}
    \end{equation}

The bound in (\ref{eq:mv_bounds_bref}) is given in terms of the TV divergence; however, the KL divergence is more commonly used in practice. The relationship between the TV divergence and KL divergence is given by Pinsker’s inequality \citep{Tsybakov2009introduction}, which demonstrates that for any two distributions $p$ and $q$: $D_{\text{TV}}(p\|q)\le \sqrt{D_{\text{KL}}(p\|q)/2}$. Then,
\begin{align*}
    \mathbb{E}_{s\sim \pi^{\bm\mu}}[D_{\text{TV}}(\bm\mu'(\cdot|s) \| \bm\mu(\cdot|s))] 
    & \le \mathbb{E}_{s\sim \pi^{\bm\mu}} \Big[\sqrt{D_{\text{KL}}(\bm\mu'(\cdot|s) \| \bm\mu(\cdot|s))/2} \Big] \\
    & \le \sqrt{\mathbb{E}_{s\sim \pi^{\bm\mu}}[D_{\text{KL}}(\bm\mu'(\cdot|s) \| \bm\mu(\cdot|s))]/2} ,
\end{align*}
where the second inequality comes from Jensen’s inequality. Substituting the result into (\ref{eq:mv_bounds_bref}) and giving an ordered subset $i_{1:N}$, we have
\begin{align*}
       J^{\bm\mu'} - J^{\bm\mu} 
       &\ge \mathcal{L}_{f}^{\bm\mu}(\bm\mu') - 2(\kappa^*-1)\epsilon_f [ \mathbb{E}_{s\sim \pi^{\bm\mu}} D_{\text{TV}}(\bm\mu'(\cdot|s) \| \bm\mu(\cdot|s))] \\
       &\ge \mathcal{L}_{f}^{\bm\mu}(\bm\mu') - (\kappa^*-1)\epsilon_f 
       \sqrt{2\mathbb{E}_{s\sim \pi^{\bm\mu}}[D_{\text{KL}}(\bm\mu'(\cdot|s) \| \bm\mu(\cdot|s))]} \\
       & \ge \mathcal{L}_{f}^{\bm\mu}(\bm\mu') - (\kappa^*-1)\epsilon_f 
       \sqrt{2\mathbb{E}_{s\sim \pi^{\bm\mu}} \Big[ \sum\limits_{i=1}^N D_{\text{KL}}(\bm\mu'(\cdot|s) \| \bm\mu(\cdot|s))} \Big] \\
       & \ge \mathcal{L}_{f}^{\bm\mu}(\bm\mu') - (\kappa^*-1)\epsilon_f
       \sum\limits_{i=1}^N \Big[ \sqrt{2\mathbb{E}_{s\sim \pi^{\bm\mu}} D_{\text{KL}}(\bm\mu'(\cdot|s) \| \bm\mu(\cdot|s)) } \Big] \\
       & = \sum\limits_{h=1}^N \Big\{  \mathcal{L}^{\bm\mu}_{i_{1:h}}(\bm\mu'_{i_{1:h-1}},\mu'_{i_h}) - (\kappa^*-1)\epsilon_f  \sqrt{2\mathbb{E}_{s\sim \pi^{\bm\mu}} D_{\text{KL}}(\mu'_{i_h}(\cdot|s) \| \mu_{i_h}(\cdot|s)) }  \Big\} .    
\end{align*}
The third inequality follows Lemma 8 in \cite{kuba2022trust}, the fourth inequality follows the Cauchy-Schwarz inequality and the last equality follows the results of Definition~\ref{def:ma_surrogate} and Equation~(\ref{eq:ad_decompostion}). Then, we complete this proof.
\endproof

\section{Details for Solving the Optimization Problem}\label{sec:appd_implementation details}
To solve the optimization problem (\ref{eq:trust_indiv}), we follow the steps of the standard trust region optimization method outlined in \cite{schulman2015trust} and \cite{zhong2024heterogeneous}. Specifically, the objective function and KL constraint in (\ref{eq:trust_indiv}) are approximated linearly and quadratically, respectively. Then, the solution to (\ref{eq:trust_indiv}) has the following closed-form expression
\begin{equation*}
    \theta_{i_h}^{(k+1)}=\theta_{i_h}^{(k)} + \alpha_{i_h} \sqrt{\frac{2\epsilon}{\bm{g_{i_h}^{(k)}}(\bm{H}_{i_h}^{(k)})^{-1} \bm{g_{i_h}^{(k)}}}} (\bm{H}_{i_h}^{(k)})^{-1} \bm{g_{i_h}^{(k)}},
\end{equation*}
where $\bm{H}_{i_h}^{(k)}=\nabla_{\theta_{i_h}}^2 \mathbb{E}_{s\sim \pi^{{\theta^{(k)}}}} \left[D_{\text{KL}}(\mu^{\theta_{i_h}^{(k)}}(\cdot|s), \mu^{\theta_{i_h}}(\cdot|s))\big|_{\theta_{i_h}=\theta_{i_h}^{(k)}}\right]$ represents the Hessian of the expected KL divergence, $\bm{g}_{i_h}^{(k)}$ denotes the gradient of the objective function in (\ref{eq:trust_indiv}), $\alpha_{i_h}<1$  is a positive coefficient determined via backtracking line search, and the product $(\bm{H}_{i_h}^{(k)})^{-1} \bm{g_{i_h}^{(k)}}$ can be efficiently computed using the conjugate gradient algorithm.

\section{Experimental Settings}
\subsection{Variable Definitions and Parameter Settings}
\label{sec:app_setting_details}
The energy management policies of agents correspond to a time scale of hours. All the continuous variables are discretized properly. The MV-TSG model is given as follows:
\begin{itemize}
    \item System state: $s_t=(G_{1,t},L_{1,t},E_{1,t},\ldots,G_{N,t},L_{N,t},E_{N,t})$, where $G_{i,t}$ denotes the generated power of microgrid $i$ at time $t$, $L_{i,t}$ denotes  demand load power, and $E_{i,t}$ denotes storage energy level.
    \item Action: $a_{i,t}=(c_{i,t},v_{i,t})$, where $c_{i,t}$ denotes the discharging power of the storage at time $t$ ($c_{i,t}<0$ means the charging power), $v_{i,t}$ denotes the power abandoned at time $t$ and $0\le v_{i,t} \le G_{i,t}$. 
    \item Transition function: Since the power generated by renewable energy generators and consumed by demand load units depend on various random factors such as climate conditions, $G_{i,t}$ and $L_{i,t}$ are non-negative random variables, and their dynamics are modeled using Markov chains. The transition of storage energy level is given by $E_{i,t+1}=E_{i,t}-c_{i,t}$.
    \item Constraints: The storage discharging power  $c_{i,t}$ must satisfy following capacity constraints: (1) $ C^{\text{min}}_i \le c_{i,t} \le C^{\text{max}}_i$, where $C^{\text{min}}_i$ and $C^{\text{max}}_i$ represent the maximum charging and discharging power, respectively. (2)  $E_{i,t} - E^{\text{max}}_i\le c_{i,t} \le E_{i,t}$, where $E^{\text{max}}_i$ represents the maximum capacity of the storage unit of microgrid $i$.
    \item Reward: $r_t= \sum\limits_{i=1}^N (G_{i,t} -L_{i,t} + c_{i,t}-v_{i,t})$, which represents the total exchanged power between the MMS and the main grid.
\end{itemize}

For simplicity, we assume that renewable energy generators, demand load units, and storage units among all microgrids share the same specifications and their stochastic characteristics are independent and identically distributed. Specifically, we focus on the wind turbine as a renewable energy generator. For each microgrid, the storage maximum capacity is $E^{\text{max}}=5$, and the maximum charging and discharging power are $C^{\text{min}}=-2$ and $C^{\text{max}}=2$. The states of wind power, demand load, and storage energy level are all divided into six states, respectively. The state details of these facilities are presented in Table~\ref{tab:state_para}.
Table~\ref{tab:action_c} illustrates the actions and their corresponding operations of storage units.

\begin{table}[htbp]
    \centering
    \caption{{\small States of different facilities in microgrids}}\label{tab:state_para}
    \begin{tabular}{ccccccc}
        \hline
        \toprule 
        State & 1 & 2 & 3 & 4 & 5 & 6\\ \hline
        Wind power/MW & 0 & 1 & 2 & 3 & 4 & 5 \\ \hline
        Demand load/MW & 0.6 & 1.2 & 1.8 & 2.4 & 3.0 & 3.6 \\ \hline
        Storage energy level/MWh & 0  & 1  & 2  & 3 & 4 & 5 \\ \hline
    \end{tabular}
\end{table}

\begin{table}[htbp]
    \centering
    \caption{{\small Scheduling actions of storage units}}\label{tab:action_c}
    \begin{tabular}{cccccc}
        \hline
        \toprule 
        Action $c$  & $-2$ & $-1$ & 0 & 1 & 2 \\ \hline
        Storage discharging power/MW & $-2$ & $-1$ & 0 & $+1$ & $+2$ \\ \hline
    \end{tabular}
\end{table}

The transition probability matrix $P_g$ in (\ref{eq:P_g}) for wind power states is estimated from statistical analysis. The real wind speed data used for this estimation is provided by the Measurement and Instrumentation Data Center at the National Renewable Energy Laboratory, which has been collecting data since 1996.

\begin{equation}\label{eq:P_g}
\bm{P}_g= \left(
\begin{matrix}
0.53 & 0.18 & 0.19 & 0.04 & 0.01 & 0.05\\
0.51 & 0.08 & 0.20 & 0.08 & 0.02 & 0.11\\
0.35 & 0.11 & 0.19 & 0.11 & 0.03 & 0.21\\
0.27 & 0.15 & 0.15 & 0.14 & 0.03 & 0.26\\
0.14 & 0.11 & 0.13 & 0.15 & 0.05 & 0.42\\
0.09 & 0.03 & 0.06 & 0.06 & 0.03 & 0.73
\end{matrix}
\right).
\end{equation}

The transition probability matrix $P_d$ in (\ref{eq:P_d}) for demand load unit states is estimated based on data from a public database established by an independent electricity system operator \citep{su2010microgrid}.

\begin{equation}\label{eq:P_d}
\bm{P}_d= \left(
\begin{matrix}
0.96 & 0.04 & 0.00 & 0.00 & 0.00 & 0.00\\
0.12 & 0.74 & 0.14 & 0.00 & 0.00 & 0.00\\
0.00 & 0.14 & 0.66 & 0.19 & 0.01 & 0.00\\
0.00 & 0.00 & 0.06 & 0.77 & 0.16 & 0.01\\
0.00 & 0.00 & 0.01 & 0.22 & 0.61 & 0.16\\
0.00 & 0.00 & 0.00 & 0.01 & 0.16 & 0.83
\end{matrix}
\right).
\end{equation}
\subsection{Hyper-parameters of MV-MATRPO}
\label{sec:appd_hyperparameter}
Table~\ref{tab:hyper} describes common hyper-parameters in MV-MVTRPO. `Total steps' denotes the number of training steps. `Number of envs' specifies the number of environments collecting data in parallel, which also equals the number of trajectories added to the buffer. `Episode length' refers to the length of each trajectory. `Number of mini-batch' indicates how many mini-batches the data batch is split into. The action/critic networks adopt a multi-layer perceptron (MLP), with a hidden size of 64 and one hidden layer. The Rectified Linear Unit (Relu) is used as the activation function. `Optimization epochs' indicates the number of iterations over the entire training dataset during the training phase. 
\begin{table}[H]
\caption{Hyper-parameters sheet}
\setstretch{0.9}
\centering
\begin{tabular}{lr}
\toprule
Hyper-parameter & Value \\
\midrule
 Total steps & 2e7 \\
 Number of envs & 8 \\
 Episode length $T$ & 1000 \\
 Number of mini-batch & 40 \\
 Actor/critic network & MLP \\
 Network hidden sizes  & 64 \\
 Hidden layer & 1 \\
 Activation function & Relu \\
 Optimizer & Adam \\
 Network learning rate & 5e-3 \\
 Optimization epochs & 5 \\
 Max grad norm & 0.5 \\
 GAE parameter $\lambda$ & 0.95 \\
 Learning rate of average performance $\alpha$ & 0.1 \\
 Average value constraint coefficient in AVC & 0.01 \\
\bottomrule
\end{tabular}
\label{tab:hyper}
\end{table}

\end{document}